
\magnification=\magstep1


\catcode`\@=11


\message{Loading jyTeX fonts...}



\font\vptrm=cmr5
\font\vptmit=cmmi5
\font\vptsy=cmsy5
\font\vptbf=cmbx5

\skewchar\vptmit='177 \skewchar\vptsy='60
\fontdimen16 \vptsy=\the\fontdimen17 \vptsy

\def\vpt{\ifmmode\err@badsizechange\else
     \@mathfontinit
     \textfont0=\vptrm  \scriptfont0=\vptrm  \scriptscriptfont0=\vptrm
     \textfont1=\vptmit \scriptfont1=\vptmit \scriptscriptfont1=\vptmit
     \textfont2=\vptsy  \scriptfont2=\vptsy  \scriptscriptfont2=\vptsy
     \textfont3=\xptex  \scriptfont3=\xptex  \scriptscriptfont3=\xptex
     \textfont\bffam=\vptbf
     \scriptfont\bffam=\vptbf
     \scriptscriptfont\bffam=\vptbf
     \@fontstyleinit
     \def\rm{\vptrm\fam=\z@}%
     \def\bf{\vptbf\fam=\bffam}%
     \def\oldstyle{\vptmit\fam=\@ne}%
     \rm\fi}


\font\viptrm=cmr6
\font\viptmit=cmmi6
\font\viptsy=cmsy6
\font\viptbf=cmbx6

\skewchar\viptmit='177 \skewchar\viptsy='60
\fontdimen16 \viptsy=\the\fontdimen17 \viptsy

\def\vipt{\ifmmode\err@badsizechange\else
     \@mathfontinit
     \textfont0=\viptrm  \scriptfont0=\vptrm  \scriptscriptfont0=\vptrm
     \textfont1=\viptmit \scriptfont1=\vptmit \scriptscriptfont1=\vptmit
     \textfont2=\viptsy  \scriptfont2=\vptsy  \scriptscriptfont2=\vptsy
     \textfont3=\xptex   \scriptfont3=\xptex  \scriptscriptfont3=\xptex
     \textfont\bffam=\viptbf
     \scriptfont\bffam=\vptbf
     \scriptscriptfont\bffam=\vptbf
     \@fontstyleinit
     \def\rm{\viptrm\fam=\z@}%
     \def\bf{\viptbf\fam=\bffam}%
     \def\oldstyle{\viptmit\fam=\@ne}%
     \rm\fi}


\font\viiptrm=cmr7
\font\viiptmit=cmmi7
\font\viiptsy=cmsy7
\font\viiptit=cmti7
\font\viiptbf=cmbx7

\skewchar\viiptmit='177 \skewchar\viiptsy='60
\fontdimen16 \viiptsy=\the\fontdimen17 \viiptsy

\def\viipt{\ifmmode\err@badsizechange\else
     \@mathfontinit
     \textfont0=\viiptrm  \scriptfont0=\vptrm  \scriptscriptfont0=\vptrm
     \textfont1=\viiptmit \scriptfont1=\vptmit \scriptscriptfont1=\vptmit
     \textfont2=\viiptsy  \scriptfont2=\vptsy  \scriptscriptfont2=\vptsy
     \textfont3=\xptex    \scriptfont3=\xptex  \scriptscriptfont3=\xptex
     \textfont\itfam=\viiptit
     \scriptfont\itfam=\viiptit
     \scriptscriptfont\itfam=\viiptit
     \textfont\bffam=\viiptbf
     \scriptfont\bffam=\vptbf
     \scriptscriptfont\bffam=\vptbf
     \@fontstyleinit
     \def\rm{\viiptrm\fam=\z@}%
     \def\it{\viiptit\fam=\itfam}%
     \def\bf{\viiptbf\fam=\bffam}%
     \def\oldstyle{\viiptmit\fam=\@ne}%
     \rm\fi}


\font\viiiptrm=cmr8
\font\viiiptmit=cmmi8
\font\viiiptsy=cmsy8
\font\viiiptit=cmti8
\font\viiiptbf=cmbx8

\skewchar\viiiptmit='177 \skewchar\viiiptsy='60
\fontdimen16 \viiiptsy=\the\fontdimen17 \viiiptsy

\def\viiipt{\ifmmode\err@badsizechange\else
     \@mathfontinit
     \textfont0=\viiiptrm  \scriptfont0=\viptrm  \scriptscriptfont0=\vptrm
     \textfont1=\viiiptmit \scriptfont1=\viptmit \scriptscriptfont1=\vptmit
     \textfont2=\viiiptsy  \scriptfont2=\viptsy  \scriptscriptfont2=\vptsy
     \textfont3=\xptex     \scriptfont3=\xptex   \scriptscriptfont3=\xptex
     \textfont\itfam=\viiiptit
     \scriptfont\itfam=\viiptit
     \scriptscriptfont\itfam=\viiptit
     \textfont\bffam=\viiiptbf
     \scriptfont\bffam=\viptbf
     \scriptscriptfont\bffam=\vptbf
     \@fontstyleinit
     \def\rm{\viiiptrm\fam=\z@}%
     \def\it{\viiiptit\fam=\itfam}%
     \def\bf{\viiiptbf\fam=\bffam}%
     \def\oldstyle{\viiiptmit\fam=\@ne}%
     \rm\fi}


\def\getixpt{%
     \font\ixptrm=cmr9
     \font\ixptmit=cmmi9
     \font\ixptsy=cmsy9
     \font\ixptit=cmti9
     \font\ixptbf=cmbx9
     \skewchar\ixptmit='177 \skewchar\ixptsy='60
     \fontdimen16 \ixptsy=\the\fontdimen17 \ixptsy}

\def\ixpt{\ifmmode\err@badsizechange\else
     \@mathfontinit
     \textfont0=\ixptrm  \scriptfont0=\viiptrm  \scriptscriptfont0=\vptrm
     \textfont1=\ixptmit \scriptfont1=\viiptmit \scriptscriptfont1=\vptmit
     \textfont2=\ixptsy  \scriptfont2=\viiptsy  \scriptscriptfont2=\vptsy
     \textfont3=\xptex   \scriptfont3=\xptex    \scriptscriptfont3=\xptex
     \textfont\itfam=\ixptit
     \scriptfont\itfam=\viiptit
     \scriptscriptfont\itfam=\viiptit
     \textfont\bffam=\ixptbf
     \scriptfont\bffam=\viiptbf
     \scriptscriptfont\bffam=\vptbf
     \@fontstyleinit
     \def\rm{\ixptrm\fam=\z@}%
     \def\it{\ixptit\fam=\itfam}%
     \def\bf{\ixptbf\fam=\bffam}%
     \def\oldstyle{\ixptmit\fam=\@ne}%
     \rm\fi}


\font\xptrm=cmr10
\font\xptmit=cmmi10
\font\xptsy=cmsy10
\font\xptex=cmex10
\font\xptit=cmti10
\font\xptsl=cmsl10
\font\xptbf=cmbx10
\font\xpttt=cmtt10
\font\xptss=cmss10
\font\xptsc=cmcsc10
\font\xptbfs=cmb10
\font\xptbmit=cmmib10

\skewchar\xptmit='177 \skewchar\xptbmit='177 \skewchar\xptsy='60
\fontdimen16 \xptsy=\the\fontdimen17 \xptsy

\def\xpt{\ifmmode\err@badsizechange\else
     \@mathfontinit
     \textfont0=\xptrm  \scriptfont0=\viiptrm  \scriptscriptfont0=\vptrm
     \textfont1=\xptmit \scriptfont1=\viiptmit \scriptscriptfont1=\vptmit
     \textfont2=\xptsy  \scriptfont2=\viiptsy  \scriptscriptfont2=\vptsy
     \textfont3=\xptex  \scriptfont3=\xptex    \scriptscriptfont3=\xptex
     \textfont\itfam=\xptit
     \scriptfont\itfam=\viiptit
     \scriptscriptfont\itfam=\viiptit
     \textfont\bffam=\xptbf
     \scriptfont\bffam=\viiptbf
     \scriptscriptfont\bffam=\vptbf
     \textfont\bfsfam=\xptbfs
     \scriptfont\bfsfam=\viiptbf
     \scriptscriptfont\bfsfam=\vptbf
     \textfont\bmitfam=\xptbmit
     \scriptfont\bmitfam=\viiptmit
     \scriptscriptfont\bmitfam=\vptmit
     \@fontstyleinit
     \def\rm{\xptrm\fam=\z@}%
     \def\it{\xptit\fam=\itfam}%
     \def\sl{\xptsl}%
     \def\bf{\xptbf\fam=\bffam}%
     \def\tt{\xpttt}%
     \def\ss{\xptss}%
     \def\sc{\xptsc}%
     \def\bfs{\xptbfs\fam=\bfsfam}%
     \def\bmit{\fam=\bmitfam}%
     \def\oldstyle{\xptmit\fam=\@ne}%
     \rm\fi}


\def\getxipt{%
     \font\xiptrm=cmr10  scaled\magstephalf
     \font\xiptmit=cmmi10 scaled\magstephalf
     \font\xiptsy=cmsy10 scaled\magstephalf
     \font\xiptex=cmex10 scaled\magstephalf
     \font\xiptit=cmti10 scaled\magstephalf
     \font\xiptsl=cmsl10 scaled\magstephalf
     \font\xiptbf=cmbx10 scaled\magstephalf
     \font\xipttt=cmtt10 scaled\magstephalf
     \font\xiptss=cmss10 scaled\magstephalf
     \skewchar\xiptmit='177 \skewchar\xiptsy='60
     \fontdimen16 \xiptsy=\the\fontdimen17 \xiptsy}

\def\xipt{\ifmmode\err@badsizechange\else
     \@mathfontinit
     \textfont0=\xiptrm  \scriptfont0=\viiiptrm  \scriptscriptfont0=\viptrm
     \textfont1=\xiptmit \scriptfont1=\viiiptmit \scriptscriptfont1=\viptmit
     \textfont2=\xiptsy  \scriptfont2=\viiiptsy  \scriptscriptfont2=\viptsy
     \textfont3=\xiptex  \scriptfont3=\xptex     \scriptscriptfont3=\xptex
     \textfont\itfam=\xiptit
     \scriptfont\itfam=\viiiptit
     \scriptscriptfont\itfam=\viiptit
     \textfont\bffam=\xiptbf
     \scriptfont\bffam=\viiiptbf
     \scriptscriptfont\bffam=\viptbf
     \@fontstyleinit
     \def\rm{\xiptrm\fam=\z@}%
     \def\it{\xiptit\fam=\itfam}%
     \def\sl{\xiptsl}%
     \def\bf{\xiptbf\fam=\bffam}%
     \def\tt{\xipttt}%
     \def\ss{\xiptss}%
     \def\oldstyle{\xiptmit\fam=\@ne}%
     \rm\fi}


\font\xiiptrm=cmr12
\font\xiiptmit=cmmi12
\font\xiiptsy=cmsy10  scaled\magstep1
\font\xiiptex=cmex10  scaled\magstep1
\font\xiiptit=cmti12
\font\xiiptsl=cmsl12
\font\xiiptbf=cmbx12
\font\xiipttt=cmtt12
\font\xiiptss=cmss12
\font\xiiptsc=cmcsc10 scaled\magstep1
\font\xiiptbfs=cmb10  scaled\magstep1
\font\xiiptbmit=cmmib10 scaled\magstep1

\skewchar\xiiptmit='177 \skewchar\xiiptbmit='177 \skewchar\xiiptsy='60
\fontdimen16 \xiiptsy=\the\fontdimen17 \xiiptsy

\def\xiipt{\ifmmode\err@badsizechange\else
     \@mathfontinit
     \textfont0=\xiiptrm  \scriptfont0=\viiiptrm  \scriptscriptfont0=\viptrm
     \textfont1=\xiiptmit \scriptfont1=\viiiptmit \scriptscriptfont1=\viptmit
     \textfont2=\xiiptsy  \scriptfont2=\viiiptsy  \scriptscriptfont2=\viptsy
     \textfont3=\xiiptex  \scriptfont3=\xptex     \scriptscriptfont3=\xptex
     \textfont\itfam=\xiiptit
     \scriptfont\itfam=\viiiptit
     \scriptscriptfont\itfam=\viiptit
     \textfont\bffam=\xiiptbf
     \scriptfont\bffam=\viiiptbf
     \scriptscriptfont\bffam=\viptbf
     \textfont\bfsfam=\xiiptbfs
     \scriptfont\bfsfam=\viiiptbf
     \scriptscriptfont\bfsfam=\viptbf
     \textfont\bmitfam=\xiiptbmit
     \scriptfont\bmitfam=\viiiptmit
     \scriptscriptfont\bmitfam=\viptmit
     \@fontstyleinit
     \def\rm{\xiiptrm\fam=\z@}%
     \def\it{\xiiptit\fam=\itfam}%
     \def\sl{\xiiptsl}%
     \def\bf{\xiiptbf\fam=\bffam}%
     \def\tt{\xiipttt}%
     \def\ss{\xiiptss}%
     \def\sc{\xiiptsc}%
     \def\bfs{\xiiptbfs\fam=\bfsfam}%
     \def\bmit{\fam=\bmitfam}%
     \def\oldstyle{\xiiptmit\fam=\@ne}%
     \rm\fi}


\def\getxiiipt{%
     \font\xiiiptrm=cmr12  scaled\magstephalf
     \font\xiiiptmit=cmmi12 scaled\magstephalf
     \font\xiiiptsy=cmsy9  scaled\magstep2
     \font\xiiiptit=cmti12 scaled\magstephalf
     \font\xiiiptsl=cmsl12 scaled\magstephalf
     \font\xiiiptbf=cmbx12 scaled\magstephalf
     \font\xiiipttt=cmtt12 scaled\magstephalf
     \font\xiiiptss=cmss12 scaled\magstephalf
     \skewchar\xiiiptmit='177 \skewchar\xiiiptsy='60
     \fontdimen16 \xiiiptsy=\the\fontdimen17 \xiiiptsy}

\def\xiiipt{\ifmmode\err@badsizechange\else
     \@mathfontinit
     \textfont0=\xiiiptrm  \scriptfont0=\xptrm  \scriptscriptfont0=\viiptrm
     \textfont1=\xiiiptmit \scriptfont1=\xptmit \scriptscriptfont1=\viiptmit
     \textfont2=\xiiiptsy  \scriptfont2=\xptsy  \scriptscriptfont2=\viiptsy
     \textfont3=\xivptex   \scriptfont3=\xptex  \scriptscriptfont3=\xptex
     \textfont\itfam=\xiiiptit
     \scriptfont\itfam=\xptit
     \scriptscriptfont\itfam=\viiptit
     \textfont\bffam=\xiiiptbf
     \scriptfont\bffam=\xptbf
     \scriptscriptfont\bffam=\viiptbf
     \@fontstyleinit
     \def\rm{\xiiiptrm\fam=\z@}%
     \def\it{\xiiiptit\fam=\itfam}%
     \def\sl{\xiiiptsl}%
     \def\bf{\xiiiptbf\fam=\bffam}%
     \def\tt{\xiiipttt}%
     \def\ss{\xiiiptss}%
     \def\oldstyle{\xiiiptmit\fam=\@ne}%
     \rm\fi}


\font\xivptrm=cmr12   scaled\magstep1
\font\xivptmit=cmmi12  scaled\magstep1
\font\xivptsy=cmsy10  scaled\magstep2
\font\xivptex=cmex10  scaled\magstep2
\font\xivptit=cmti12  scaled\magstep1
\font\xivptsl=cmsl12  scaled\magstep1
\font\xivptbf=cmbx12  scaled\magstep1
\font\xivpttt=cmtt12  scaled\magstep1
\font\xivptss=cmss12  scaled\magstep1
\font\xivptsc=cmcsc10 scaled\magstep2
\font\xivptbfs=cmb10  scaled\magstep2
\font\xivptbmit=cmmib10 scaled\magstep2

\skewchar\xivptmit='177 \skewchar\xivptbmit='177 \skewchar\xivptsy='60
\fontdimen16 \xivptsy=\the\fontdimen17 \xivptsy

\def\xivpt{\ifmmode\err@badsizechange\else
     \@mathfontinit
     \textfont0=\xivptrm  \scriptfont0=\xptrm  \scriptscriptfont0=\viiptrm
     \textfont1=\xivptmit \scriptfont1=\xptmit \scriptscriptfont1=\viiptmit
     \textfont2=\xivptsy  \scriptfont2=\xptsy  \scriptscriptfont2=\viiptsy
     \textfont3=\xivptex  \scriptfont3=\xptex  \scriptscriptfont3=\xptex
     \textfont\itfam=\xivptit
     \scriptfont\itfam=\xptit
     \scriptscriptfont\itfam=\viiptit
     \textfont\bffam=\xivptbf
     \scriptfont\bffam=\xptbf
     \scriptscriptfont\bffam=\viiptbf
     \textfont\bfsfam=\xivptbfs
     \scriptfont\bfsfam=\xptbfs
     \scriptscriptfont\bfsfam=\viiptbf
     \textfont\bmitfam=\xivptbmit
     \scriptfont\bmitfam=\xptbmit
     \scriptscriptfont\bmitfam=\viiptmit
     \@fontstyleinit
     \def\rm{\xivptrm\fam=\z@}%
     \def\it{\xivptit\fam=\itfam}%
     \def\sl{\xivptsl}%
     \def\bf{\xivptbf\fam=\bffam}%
     \def\tt{\xivpttt}%
     \def\ss{\xivptss}%
     \def\sc{\xivptsc}%
     \def\bfs{\xivptbfs\fam=\bfsfam}%
     \def\bmit{\fam=\bmitfam}%
     \def\oldstyle{\xivptmit\fam=\@ne}%
     \rm\fi}


\font\xviiptrm=cmr17
\font\xviiptmit=cmmi12 scaled\magstep2
\font\xviiptsy=cmsy10 scaled\magstep3
\font\xviiptex=cmex10 scaled\magstep3
\font\xviiptit=cmti12 scaled\magstep2
\font\xviiptbf=cmbx12 scaled\magstep2
\font\xviiptbfs=cmb10 scaled\magstep3

\skewchar\xviiptmit='177 \skewchar\xviiptsy='60
\fontdimen16 \xviiptsy=\the\fontdimen17 \xviiptsy

\def\xviipt{\ifmmode\err@badsizechange\else
     \@mathfontinit
     \textfont0=\xviiptrm  \scriptfont0=\xiiptrm  \scriptscriptfont0=\viiiptrm
     \textfont1=\xviiptmit \scriptfont1=\xiiptmit \scriptscriptfont1=\viiiptmit
     \textfont2=\xviiptsy  \scriptfont2=\xiiptsy  \scriptscriptfont2=\viiiptsy
     \textfont3=\xviiptex  \scriptfont3=\xiiptex  \scriptscriptfont3=\xptex
     \textfont\itfam=\xviiptit
     \scriptfont\itfam=\xiiptit
     \scriptscriptfont\itfam=\viiiptit
     \textfont\bffam=\xviiptbf
     \scriptfont\bffam=\xiiptbf
     \scriptscriptfont\bffam=\viiiptbf
     \textfont\bfsfam=\xviiptbfs
     \scriptfont\bfsfam=\xiiptbfs
     \scriptscriptfont\bfsfam=\viiiptbf
     \@fontstyleinit
     \def\rm{\xviiptrm\fam=\z@}%
     \def\it{\xviiptit\fam=\itfam}%
     \def\bf{\xviiptbf\fam=\bffam}%
     \def\bfs{\xviiptbfs\fam=\bfsfam}%
     \def\oldstyle{\xviiptmit\fam=\@ne}%
     \rm\fi}


\font\xxiptrm=cmr17  scaled\magstep1


\def\xxipt{\ifmmode\err@badsizechange\else
     \@mathfontinit
     \@fontstyleinit
     \def\rm{\xxiptrm\fam=\z@}%
     \rm\fi}


\font\xxvptrm=cmr17  scaled\magstep2


\def\xxvpt{\ifmmode\err@badsizechange\else
     \@mathfontinit
     \@fontstyleinit
     \def\rm{\xxvptrm\fam=\z@}%
     \rm\fi}




\message{Loading jyTeX macros...}

\message{modifications to plain.tex,}


\def\newcount{\alloc@0\count\countdef\insc@unt}
\def\newdimen{\alloc@1\dimen\dimendef\insc@unt}
\def\newskip{\alloc@2\skip\skipdef\insc@unt}
\def\newmuskip{\alloc@3\muskip\muskipdef\@cclvi}
\def\newbox{\alloc@4\box\chardef\insc@unt}
\def\newtoks{\alloc@5\toks\toksdef\@cclvi}
\def\newhelp#1#2{\newtoks#1\global#1\expandafter{\csname#2\endcsname}}
\def\newread{\alloc@6\read\chardef\sixt@@n}
\def\newwrite{\alloc@7\write\chardef\sixt@@n}
\def\newfam{\alloc@8\fam\chardef\sixt@@n}
\def\newinsert#1{\global\advance\insc@unt by\m@ne
     \ch@ck0\insc@unt\count
     \ch@ck1\insc@unt\dimen
     \ch@ck2\insc@unt\skip
     \ch@ck4\insc@unt\box
     \allocationnumber=\insc@unt
     \global\chardef#1=\allocationnumber
     \wlog{\string#1=\string\insert\the\allocationnumber}}
\def\newif#1{\count@\escapechar \escapechar\m@ne
     \expandafter\expandafter\expandafter
          \xdef\@if#1{true}{\let\noexpand#1=\noexpand\iftrue}%
     \expandafter\expandafter\expandafter
          \xdef\@if#1{false}{\let\noexpand#1=\noexpand\iffalse}%
     \global\@if#1{false}\escapechar=\count@}


\newlinechar=`\^^J
\overfullrule=0pt




\let\itfam=\undefined

\let\bffam=\undefined

\count18=3


\chardef\sharps="19


\mathchardef\alpha="710B
\mathchardef\beta="710C
\mathchardef\gamma="710D
\mathchardef\delta="710E
\mathchardef\epsilon="710F
\mathchardef\zeta="7110
\mathchardef\eta="7111
\mathchardef\theta="7112
\mathchardef\iota="7113
\mathchardef\kappa="7114
\mathchardef\lambda="7115
\mathchardef\mu="7116
\mathchardef\nu="7117
\mathchardef\xi="7118
\mathchardef\pi="7119
\mathchardef\rho="711A
\mathchardef\sigma="711B
\mathchardef\tau="711C
\mathchardef\upsilon="711D
\mathchardef\phi="711E
\mathchardef\chi="711F
\mathchardef\psi="7120
\mathchardef\omega="7121
\mathchardef\varepsilon="7122
\mathchardef\vartheta="7123
\mathchardef\varpi="7124
\mathchardef\varrho="7125
\mathchardef\varsigma="7126
\mathchardef\varphi="7127
\mathchardef\imath="717B
\mathchardef\jmath="717C
\mathchardef\ell="7160
\mathchardef\wp="717D
\mathchardef\partial="7140
\mathchardef\flat="715B
\mathchardef\natural="715C
\mathchardef\sharp="715D



\def\angle{{\vbox{\ialign{$\m@th\scriptstyle##$\crcr
     \not\mathrel{\mkern14mu}\crcr
     \noalign{\nointerlineskip}
     \mkern2.5mu\leaders\hrule height.34\rp@\hfill\mkern2.5mu\crcr}}}}
\def\vdots{\vbox{\baselineskip4\rp@ \lineskiplimit\z@
     \kern6\rp@\hbox{.}\hbox{.}\hbox{.}}}
\def\ddots{\mathinner{\mkern1mu\raise7\rp@\vbox{\kern7\rp@\hbox{.}}\mkern2mu
     \raise4\rp@\hbox{.}\mkern2mu\raise\rp@\hbox{.}\mkern1mu}}
\def\overrightarrow#1{\vbox{\ialign{##\crcr
     \rightarrowfill\crcr
     \noalign{\kern-\rp@\nointerlineskip}
     $\hfil\displaystyle{#1}\hfil$\crcr}}}
\def\overleftarrow#1{\vbox{\ialign{##\crcr
     \leftarrowfill\crcr
     \noalign{\kern-\rp@\nointerlineskip}
     $\hfil\displaystyle{#1}\hfil$\crcr}}}
\def\overbrace#1{\mathop{\vbox{\ialign{##\crcr
     \noalign{\kern3\rp@}
     \downbracefill\crcr
     \noalign{\kern3\rp@\nointerlineskip}
     $\hfil\displaystyle{#1}\hfil$\crcr}}}\limits}
\def\underbrace#1{\mathop{\vtop{\ialign{##\crcr
     $\hfil\displaystyle{#1}\hfil$\crcr
     \noalign{\kern3\rp@\nointerlineskip}
     \upbracefill\crcr
     \noalign{\kern3\rp@}}}}\limits}
\def\big#1{{\hbox{$\left#1\vbox to8.5\rp@ {}\right.\n@space$}}}
\def\Big#1{{\hbox{$\left#1\vbox to11.5\rp@ {}\right.\n@space$}}}
\def\bigg#1{{\hbox{$\left#1\vbox to14.5\rp@ {}\right.\n@space$}}}
\def\Bigg#1{{\hbox{$\left#1\vbox to17.5\rp@ {}\right.\n@space$}}}
\def\@vereq#1#2{\lower.5\rp@\vbox{\baselineskip\z@skip\lineskip-.5\rp@
     \ialign{$\m@th#1\hfil##\hfil$\crcr#2\crcr=\crcr}}}
\def\rlh@#1{\vcenter{\hbox{\ooalign{\raise2\rp@
     \hbox{$#1\rightharpoonup$}\crcr
     $#1\leftharpoondown$}}}}
\def\bordermatrix#1{\begingroup\m@th
     \setbox\z@\vbox{%
          \def\cr{\crcr\noalign{\kern2\rp@\global\let\cr\endline}}%
          \ialign{$##$\hfil\kern2\rp@\kern\p@renwd
               &\thinspace\hfil$##$\hfil&&\quad\hfil$##$\hfil\crcr
               \omit\strut\hfil\crcr
               \noalign{\kern-\baselineskip}%
               #1\crcr\omit\strut\cr}}%
     \setbox\tw@\vbox{\unvcopy\z@\global\setbox\@ne\lastbox}%
     \setbox\tw@\hbox{\unhbox\@ne\unskip\global\setbox\@ne\lastbox}%
     \setbox\tw@\hbox{$\kern\wd\@ne\kern-\p@renwd\left(\kern-\wd\@ne
          \global\setbox\@ne\vbox{\box\@ne\kern2\rp@}%
          \vcenter{\kern-\ht\@ne\unvbox\z@\kern-\baselineskip}%
          \,\right)$}%
     \null\;\vbox{\kern\ht\@ne\box\tw@}\endgroup}
\def\endinsert{\egroup
     \if@mid\dimen@\ht\z@
          \advance\dimen@\dp\z@
          \advance\dimen@12\rp@
          \advance\dimen@\pagetotal
          \ifdim\dimen@>\pagegoal\@midfalse\p@gefalse\fi
     \fi
     \if@mid\bigskip\box\z@
          \bigbreak
     \else\insert\topins{\penalty100 \splittopskip\z@skip
               \splitmaxdepth\maxdimen\floatingpenalty\z@
               \ifp@ge\dimen@\dp\z@
                    \vbox to\vsize{\unvbox\z@\kern-\dimen@}%
               \else\box\z@\nobreak\bigskip
               \fi}%
     \fi
     \endgroup}


\def\cases#1{\left\{\,\vcenter{\m@th
     \ialign{$##\hfil$&\quad##\hfil\crcr#1\crcr}}\right.}
\def\matrix#1{\null\,\vcenter{\m@th
     \ialign{\hfil$##$\hfil&&\quad\hfil$##$\hfil\crcr
          \mathstrut\crcr
          \noalign{\kern-\baselineskip}
          #1\crcr
          \mathstrut\crcr
          \noalign{\kern-\baselineskip}}}\,}


\newif\ifraggedbottom

\def\raggedbottom{\ifraggedbottom\else
     \advance\topskip by\z@ plus60pt \raggedbottomtrue\fi}%
\def\normalbottom{\ifraggedbottom
     \advance\topskip by\z@ plus-60pt \raggedbottomfalse\fi}

\message{hacks,}


\toksdef\toks@i=1
\toksdef\toks@ii=2


\def\TeX{T\kern-.1667em \lower.5ex \hbox{E}\kern-.125em X\null}
\def\jyTeX{{\leavevmode
     \raise.587ex \hbox{\it\j}\kern-.1em \lower.048ex \hbox{\it y}\kern-.12em
     \TeX}}

\let\then=\iftrue
\def\ifnoarg#1\then{\def\hack@{#1}\ifx\hack@\empty}
\def\ifundefined#1\then{%
     \expandafter\ifx\csname\expandafter\blank\string#1\endcsname\relax}
\def\useif#1\then{\csname#1\endcsname}
\def\usename#1{\csname#1\endcsname}
\def\useafter#1#2{\expandafter#1\csname#2\endcsname}

\long\def\loop#1\repeat{\def\@iterate{#1\expandafter\@iterate\fi}\@iterate
     \let\@iterate=\relax}

\let\TeXend=\end
\def\begin#1{\begingroup\def\@@blockname{#1}\usename{begin#1}}
\def\end#1{\usename{end#1}\def\hack@{#1}%
     \ifx\@@blockname\hack@
          \endgroup
     \else\err@badgroup\hack@\@@blockname
     \fi}
\def\@@blockname{}

\def\defaultoption[#1]#2{%
     \def\hack@{\ifx\hack@ii[\toks@={#2}\else\toks@={#2[#1]}\fi\the\toks@}%
     \futurelet\hack@ii\hack@}

\def\markup#1{\let\@@marksf=\empty
     \ifhmode\edef\@@marksf{\spacefactor=\the\spacefactor\relax}\/\fi
     ${}^{\hbox{\subscriptfonts#1}}$\@@marksf}


\newtoks\shortyear
\newtoks\militaryhour
\newtoks\standardhour
\newtoks\minute
\newtoks\amorpm

\def\settime{\count@=\time\divide\count@ by60
     \militaryhour=\expandafter{\number\count@}%
     {\multiply\count@ by-60 \advance\count@ by\time
          \xdef\hack@{\ifnum\count@<10 0\fi\number\count@}}%
     \minute=\expandafter{\hack@}%
     \ifnum\count@<12
          \amorpm={am}
     \else\amorpm={pm}
          \ifnum\count@>12 \advance\count@ by-12 \fi
     \fi
     \standardhour=\expandafter{\number\count@}%
     \def\hack@19##1##2{\shortyear={##1##2}}%
          \expandafter\hack@\the\year}

\def\monthword#1{%
     \ifcase#1
          $\bullet$\err@badcountervalue{monthword}%
          \or January\or February\or March\or April\or May\or June%
          \or July\or August\or September\or October\or November\or December%
     \else$\bullet$\err@badcountervalue{monthword}%
     \fi}

\def\monthabbr#1{%
     \ifcase#1
          $\bullet$\err@badcountervalue{monthabbr}%
          \or Jan\or Feb\or Mar\or Apr\or May\or Jun%
          \or Jul\or Aug\or Sep\or Oct\or Nov\or Dec%
     \else$\bullet$\err@badcountervalue{monthabbr}%
     \fi}

\def\militarytime{\the\militaryhour:\the\minute}
\def\standardtime{\the\standardhour:\the\minute}


\def\@setnumstyle#1#2{\expandafter\global\expandafter\expandafter
     \expandafter\let\expandafter\expandafter
     \csname @\expandafter\blank\string#1style\endcsname
     \csname#2\endcsname}
\def\numstyle#1{\usename{@\expandafter\blank\string#1style}#1}
\def\ifblank#1\then{\useafter\ifx{@\expandafter\blank\string#1}\blank}

\def\blank#1{}

\def\Roman#1{\expandafter\uppercase\expandafter{\romannumeral#1}}
\def\alphabetic#1{%
     \ifcase#1
          $\bullet$\err@badcountervalue{alphabetic}%
          \or a\or b\or c\or d\or e\or f\or g\or h\or i\or j\or k\or l\or m%
          \or n\or o\or p\or q\or r\or s\or t\or u\or v\or w\or x\or y\or z%
     \else$\bullet$\err@badcountervalue{alphabetic}%
     \fi}
\def\Alphabetic#1{\expandafter\uppercase\expandafter{\alphabetic{#1}}}
\def\symbols#1{%
     \ifcase#1
          $\bullet$\err@badcountervalue{symbols}%
          \or*\or\dag\or\ddag\or\S\or$\|$%
          \or**\or\dag\dag\or\ddag\ddag\or\S\S\or$\|\|$%
     \else$\bullet$\err@badcountervalue{symbols}%
     \fi}


\catcode`\^^?=13 \def^^?{\relax}

\def\trimleading#1\to#2{\edef#2{#1}%
     \expandafter\@trimleading\expandafter#2#2^^?^^?}
\def\@trimleading#1#2#3^^?{\ifx#2^^?\def#1{}\else\def#1{#2#3}\fi}

\def\trimtrailing#1\to#2{\edef#2{#1}%
     \expandafter\@trimtrailing\expandafter#2#2^^? ^^?\relax}
\def\@trimtrailing#1#2 ^^?#3{\ifx#3\relax\toks@={}%
     \else\def#1{#2}\toks@={\trimtrailing#1\to#1}\fi
     \the\toks@}

\def\trim#1\to#2{\trimleading#1\to#2\trimtrailing#2\to#2}

\catcode`\^^?=15


\long\def\additemL#1\to#2{\toks@={\^^\{#1}}\toks@ii=\expandafter{#2}%
     \xdef#2{\the\toks@\the\toks@ii}}

\long\def\additemR#1\to#2{\toks@={\^^\{#1}}\toks@ii=\expandafter{#2}%
     \xdef#2{\the\toks@ii\the\toks@}}

\def\getitemL#1\to#2{\expandafter\@getitemL#1\hack@#1#2}
\def\@getitemL\^^\#1#2\hack@#3#4{\def#4{#1}\def#3{#2}}

\message{font macros,}


\newdimen\rp@
\newcount\@@sizeindex \@@sizeindex=0
\newcount\@@factori
\newcount\@@factorii
\newcount\@@factoriii
\newcount\@@factoriv

\countdef\maxfam=18
\newfam\itfam
\newfam\bffam
\newfam\bfsfam
\newfam\bmitfam

\def\@mathfontinit{\count@=4
     \loop\textfont\count@=\nullfont
          \scriptfont\count@=\nullfont
          \scriptscriptfont\count@=\nullfont
          \ifnum\count@<\maxfam\advance\count@ by\@ne
     \repeat}

\def\@fontstyleinit{%
     \def\it{\err@fontnotavailable\it}%
     \def\bf{\err@fontnotavailable\bf}%
     \def\bfs{\err@bfstobf}%
     \def\bmit{\err@fontnotavailable\bmit}%
     \def\sc{\err@fontnotavailable\sc}%
     \def\sl{\err@sltoit}%
     \def\ss{\err@fontnotavailable\ss}%
     \def\tt{\err@fontnotavailable\tt}}

\def\@parameterinit#1{\rm\rp@=.1em \@getscaling{#1}%
     \let\^^\=\@doscaling\scalingskipslist
     \setbox\strutbox=\hbox{\vrule
          height.708\baselineskip depth.292\baselineskip width\z@}}

\def\@getfactor#1#2#3#4{\@@factori=#1 \@@factorii=#2
     \@@factoriii=#3 \@@factoriv=#4}

\def\@getscaling#1{\count@=#1 \advance\count@ by-\@@sizeindex\@@sizeindex=#1
     \ifnum\count@<0
          \let\@mulordiv=\divide
          \let\@divormul=\multiply
          \multiply\count@ by\m@ne
     \else\let\@mulordiv=\multiply
          \let\@divormul=\divide
     \fi
     \edef\@@scratcha{\ifcase\count@                {1}{1}{1}{1}\or
          {1}{7}{23}{3}\or     {2}{5}{3}{1}\or      {9}{89}{13}{1}\or
          {6}{25}{6}{1}\or     {8}{71}{14}{1}\or    {6}{25}{36}{5}\or
          {1}{7}{53}{4}\or     {12}{125}{108}{5}\or {3}{14}{53}{5}\or
          {6}{41}{17}{1}\or    {13}{31}{13}{2}\or   {9}{107}{71}{2}\or
          {11}{139}{124}{3}\or {1}{6}{43}{2}\or     {10}{107}{42}{1}\or
          {1}{5}{43}{2}\or     {5}{69}{65}{1}\or    {11}{97}{91}{2}\fi}%
     \expandafter\@getfactor\@@scratcha}

\def\@doscaling#1{\@mulordiv#1by\@@factori\@divormul#1by\@@factorii
     \@mulordiv#1by\@@factoriii\@divormul#1by\@@factoriv}


\newskip\headskip
\newskip\footskip

\def\typesize=#1pt{\count@=#1 \advance\count@ by-10
     \ifcase\count@
          \@setsizex\or\err@badtypesize\or
          \@setsizexii\or\err@badtypesize\or
          \@setsizexiv
     \else\err@badtypesize
     \fi}

\def\@setsizex{\getixpt
     \def\subsubscriptfonts{\vpt}%
          \def\subsubscriptsize{\vpt\@parameterinit{-8}}%
     \def\subscriptfonts{\viipt}\def\subscriptsize{\viipt\@parameterinit{-4}}%
     \def\footnotefonts{\viiipt}\def\footnotesize{\viiipt\@parameterinit{-2}}%
     \def\smallfonts{\ixpt}\def\smallsize{\ixpt\@parameterinit{-1}}%
     \def\normalfonts{\xpt}\def\normalsize{\xpt\@parameterinit{0}}%
     \def\bigfonts{\xiipt}\def\bigsize{\xiipt\@parameterinit{2}}%
     \def\Bigfonts{\xivpt}\def\Bigsize{\xivpt\@parameterinit{4}}%
     \def\biggfonts{\xviipt}\def\biggsize{\xviipt\@parameterinit{6}}%
     \def\Biggfonts{\xxipt}\def\Biggsize{\xxipt\@parameterinit{8}}%
     \def\tinyfonts{\vpt}\def\tinysize{\vpt\@parameterinit{-8}}%
     \def\HUGEFONTS{\xxvpt}\def\HUGESIZE{\xxvpt\@parameterinit{10}}%
     \normalsize\fixedskipslist}

\def\@setsizexii{\getxipt
     \def\subsubscriptfonts{\vipt}%
          \def\subsubscriptsize{\vipt\@parameterinit{-6}}%
     \def\subscriptfonts{\viiipt}%
          \def\subscriptsize{\viiipt\@parameterinit{-2}}%
     \def\footnotefonts{\xpt}\def\footnotesize{\xpt\@parameterinit{0}}%
     \def\smallfonts{\xipt}\def\smallsize{\xipt\@parameterinit{1}}%
     \def\normalfonts{\xiipt}\def\normalsize{\xiipt\@parameterinit{2}}%
     \def\bigfonts{\xivpt}\def\bigsize{\xivpt\@parameterinit{4}}%
     \def\Bigfonts{\xviipt}\def\Bigsize{\xviipt\@parameterinit{6}}%
     \def\biggfonts{\xxipt}\def\biggsize{\xxipt\@parameterinit{8}}%
     \def\Biggfonts{\xxvpt}\def\Biggsize{\xxvpt\@parameterinit{10}}%
     \def\tinyfonts{\vpt}\def\tinysize{\vpt\@parameterinit{-8}}%
     \def\HUGEFONTS{\xxvpt}\def\HUGESIZE{\xxvpt\@parameterinit{10}}%
     \normalsize\fixedskipslist}

\def\@setsizexiv{\getxiiipt
     \def\subsubscriptfonts{\viipt}%
          \def\subsubscriptsize{\viipt\@parameterinit{-4}}%
     \def\subscriptfonts{\xpt}\def\subscriptsize{\xpt\@parameterinit{0}}%
     \def\footnotefonts{\xiipt}\def\footnotesize{\xiipt\@parameterinit{2}}%
     \def\smallfonts{\xiiipt}\def\smallsize{\xiiipt\@parameterinit{3}}%
     \def\normalfonts{\xivpt}\def\normalsize{\xivpt\@parameterinit{4}}%
     \def\bigfonts{\xviipt}\def\bigsize{\xviipt\@parameterinit{6}}%
     \def\Bigfonts{\xxipt}\def\Bigsize{\xxipt\@parameterinit{8}}%
     \def\biggfonts{\xxvpt}\def\biggsize{\xxvpt\@parameterinit{10}}%
     \def\Biggfonts{\err@sizetoolarge\Biggfonts\HUGEFONTS}%
          \def\Biggsize{\err@sizetoolarge\Biggsize\HUGESIZE}%
     \def\tinyfonts{\vpt}\def\tinysize{\vpt\@parameterinit{-8}}%
     \def\HUGEFONTS{\xxvpt}\def\HUGESIZE{\xxvpt\@parameterinit{10}}%
     \normalsize\fixedskipslist}

\def\subsubscriptfonts{\vpt} \def\subsubscriptsize{\vpt\@parameterinit{-8}}
\def\subscriptfonts{\viipt}  \def\subscriptsize{\viipt\@parameterinit{-4}}
\def\footnotefonts{\viiipt}  \def\footnotesize{\viiipt\@parameterinit{-2}}
\def\smallfonts{\err@sizenotavailable\smallfonts}
                             \def\smallsize{\ixpt\@parameterinit{-1}}
\def\normalfonts{\xpt}       \def\normalsize{\xpt\@parameterinit{0}}
\def\bigfonts{\xiipt}        \def\bigsize{\xiipt\@parameterinit{2}}
\def\Bigfonts{\xivpt}        \def\Bigsize{\xivpt\@parameterinit{4}}
\def\biggfonts{\xviipt}      \def\biggsize{\xviipt\@parameterinit{6}}
\def\Biggfonts{\xxipt}       \def\Biggsize{\xxipt\@parameterinit{8}}
\def\tinyfonts{\vpt}         \def\tinysize{\vpt\@parameterinit{-8}}
\def\HUGEFONTS{\xxvpt}       \def\HUGESIZE{\xxvpt\@parameterinit{10}}

\message{document layout,}


\newtoks\everyoutput \everyoutput={}
\newdimen\depthofpage
\newcount\pagenum \pagenum=0

\newdimen\oddtopmargin  \newdimen\eventopmargin\newdimen\oddleftmargin
\newdimen\evenleftmargin
\newtoks\oddhead        \newtoks\evenhead
\newtoks\oddfoot        \newtoks\evenfoot

\def\topmargin{\afterassignment\@seteventop\oddtopmargin}
\def\leftmargin{\afterassignment\@setevenleft\oddleftmargin}
\def\head{\afterassignment\@setevenhead\oddhead}
\def\foot{\afterassignment\@setevenfoot\oddfoot}

\def\@seteventop{\eventopmargin=\oddtopmargin}
\def\@setevenleft{\evenleftmargin=\oddleftmargin}
\def\@setevenhead{\evenhead=\oddhead}
\def\@setevenfoot{\evenfoot=\oddfoot}

\def\pagenumstyle#1{\@setnumstyle\pagenum{#1}}

\newif\ifdraft
\def\draft{\drafttrue\leftmargin=.5in \overfullrule=5pt }

\def\outputstyle#1{\global\expandafter\let\expandafter
          \@outputstyle\csname#1output\endcsname
     \usename{#1setup}}

\output={\@outputstyle}

\def\normaloutput{\the\everyoutput
     \global\advance\pagenum by\@ne
     \ifodd\pagenum
          \voffset=\oddtopmargin \hoffset=\oddleftmargin
     \else\voffset=\eventopmargin \hoffset=\evenleftmargin
     \fi
     \advance\voffset by-1in  \advance\hoffset by-1in
     \count0=\pagenum
     \expandafter\shipout\pagebox
     \ifnum\outputpenalty>-\@MM\else\dosupereject\fi}

\newdimen\fullhsize
\newbox\leftpage
\newcount\leftpagenum
\newcount\outputpagenum \outputpagenum=0
\let\leftorright=L

\def\twoupoutput{\the\everyoutput
     \global\advance\pagenum by\@ne
     \if L\leftorright
          \global\setbox\leftpage=\leftline{\pagebox}%
          \global\leftpagenum=\pagenum
          \global\let\leftorright=R%
     \else\global\advance\outputpagenum by\@ne
          \ifodd\outputpagenum
               \voffset=\oddtopmargin \hoffset=\oddleftmargin
          \else\voffset=\eventopmargin \hoffset=\evenleftmargin
          \fi
          \advance\voffset by-1in  \advance\hoffset by-1in
          \count0=\leftpagenum \count1=\pagenum
          \shipout\vbox{\hbox to\fullhsize
               {\box\leftpage\hfil\leftline{\pagebox}}}%
          \global\let\leftorright=L%
     \fi
     \ifnum\outputpenalty>-\@MM
     \else\dosupereject
          \if R\leftorright
               \globaldefs=\@ne\head={\hfil}\foot={\hfil}\globaldefs=\z@
               \null\newpage
          \fi
     \fi}

\def\pagebox{\vbox{\makeheadline\pagebody\makefootline}}

\def\makeheadline{%
     \vbox to\z@{\baselinestretch=\@m
          \vskip\topskip\vskip-.708\baselineskip\vskip-\headskip
          \line{\vbox to\ht\strutbox{}%
               \ifodd\pagenum\the\oddhead\else\the\evenhead\fi}%
          \vss}%
     \nointerlineskip}

\def\pagebody{\vbox to\vsize{%
     \boxmaxdepth\maxdepth
     \ifvoid\topins\else\unvbox\topins\fi
     \depthofpage=\dp255
     \unvbox255
     \ifraggedbottom\kern-\depthofpage\vfil\fi
     \ifvoid\footins
     \else\vskip\skip\footins
          \footnoterule
          \unvbox\footins
          \vskip-\footnoteskip
     \fi}}

\def\makefootline{\baselineskip=\footskip
     \line{\ifodd\pagenum\the\oddfoot\else\the\evenfoot\fi}}


\newskip\abovechapterskip
\newskip\belowchapterskip
\newskip\abovesectionskip
\newskip\belowsectionskip
\newskip\abovesubsectionskip
\newskip\belowsubsectionskip

\def\chapterstyle#1{\global\expandafter\let\expandafter\@chapterstyle
     \csname#1text\endcsname}
\def\sectionstyle#1{\global\expandafter\let\expandafter\@sectionstyle
     \csname#1text\endcsname}
\def\subsectionstyle#1{\global\expandafter\let\expandafter\@subsectionstyle
     \csname#1text\endcsname}

\def\chapter#1{%
     \ifdim\lastskip=17sp \else\chapterbreak\vskip\abovechapterskip\fi
     \@chapterstyle{\ifblank\chapternumstyle\then
          \else\newchapternum=\next\chapternumformat\ \fi#1}%
     \nobreak\vskip\belowchapterskip\vskip17sp }

\def\section#1{%
     \ifdim\lastskip=17sp \else\sectionbreak\vskip\abovesectionskip\fi
     \@sectionstyle{\ifblank\sectionnumstyle\then
          \else\newsectionnum=\next\sectionnumformat\ \fi#1}%
     \nobreak\vskip\belowsectionskip\vskip17sp }

\def\subsection#1{%
     \ifdim\lastskip=17sp \else\subsectionbreak\vskip\abovesubsectionskip\fi
     \@subsectionstyle{\ifblank\subsectionnumstyle\then
          \else\newsubsectionnum=\next\subsectionnumformat\ \fi#1}%
     \nobreak\vskip\belowsubsectionskip\vskip17sp }


\let\TeXunderline=\underline
\let\TeXoverline=\overline
\def\underline#1{\relax\ifmmode\TeXunderline{#1}\else
     $\TeXunderline{\hbox{#1}}$\fi}
\def\overline#1{\relax\ifmmode\TeXoverline{#1}\else
     $\TeXoverline{\hbox{#1}}$\fi}

\def\baselinestretch{\afterassignment\@baselinestretch\count@}
\def\@baselinestretch{\baselineskip=\normalbaselineskip
     \divide\baselineskip by\@m\baselineskip=\count@\baselineskip
     \setbox\strutbox=\hbox{\vrule
          height.708\baselineskip depth.292\baselineskip width\z@}%
     \bigskipamount=\the\baselineskip
          plus.25\baselineskip minus.25\baselineskip
     \medskipamount=.5\baselineskip
          plus.125\baselineskip minus.125\baselineskip
     \smallskipamount=.25\baselineskip
          plus.0625\baselineskip minus.0625\baselineskip}

\def\\{\ifhmode\ifnum\lastpenalty=-\@M\else\hfil\penalty-\@M\fi\fi
     \ignorespaces}
\def\newpage{\vfil\break}

\def\lefttext#1{\par{\@text\leftskip=\z@\rightskip=\centering
     \noindent#1\par}}
\def\righttext#1{\par{\@text\leftskip=\centering\rightskip=\z@
     \noindent#1\par}}
\def\centertext#1{\par{\@text\leftskip=\centering\rightskip=\centering
     \noindent#1\par}}
\def\@text{\parindent=\z@ \parfillskip=\z@ \everypar={}%
     \spaceskip=.3333em \xspaceskip=.5em
     \def\\{\ifhmode\ifnum\lastpenalty=-\@M\else\penalty-\@M\fi\fi
          \ignorespaces}}

\def\beginleft{\par\@text\leftskip=\z@ \rightskip=\centering}
     
\def\beginright{\par\@text\leftskip=\centering\rightskip=\z@ }
     
\def\begincenter{\par\@text\leftskip=\centering\rightskip=\centering}

\def\beginnarrow{\defaultoption[\parindent]\@beginnarrow}
\def\@beginnarrow[#1]{\par\advance\leftskip by#1\advance\rightskip by#1}

\begingroup
\catcode`\[=1 \catcode`\{=11
\gdef\beginignore[\endgroup\bgroup
     \catcode`\e=0 \catcode`\\=12 \catcode`\{=11 \catcode`\f=12 \let\or=\relax
     \let\nd{ignor=\fi \let\}=\egroup
     \iffalse}
\endgroup

\long\def\marginnote#1{\leavevmode
     \edef\@marginsf{\spacefactor=\the\spacefactor\relax}%
     \ifdraft\strut\vadjust{%
          \hbox to\z@{\hskip\hsize\hskip.1in
               \vbox to\z@{\vskip-\dp\strutbox
                    \marginnoteformat
                    \vskip-\ht\strutbox
                    \noindent\strut#1\par
                    \vss}%
               \hss}}%
     \fi
     \@marginsf}


\newtoks\everybye \everybye={\par\vfil}
\outer\def\bye{\the\everybye
     \footnotecheck
     \prelabelcheck
     \streamcheck
     \supereject
     \TeXend}

\message{footnotes,}

\newcount\footnotenum \footnotenum=0
\newskip\footnoteskip
\let\@footnotelist=\empty

\def\footnotenumstyle#1{\@setnumstyle\footnotenum{#1}%
     \useafter\ifx{@footnotenumstyle}\symbols
          \global\let\@footup=\empty
     \else\global\let\@footup=\markup
     \fi}

\def\footnote{\footnotecheck\defaultoption[]\@footnote}
\def\@footnote[#1]{\@footnotemark[#1]\@footnotetext}

\def\footnotemark{\defaultoption[]\@footnotemark}
\def\@footnotemark[#1]{\let\@footsf=\empty
     \ifhmode\edef\@footsf{\spacefactor=\the\spacefactor\relax}\/\fi
     \ifnoarg#1\then
          \global\advance\footnotenum by\@ne
          \@footup{\footnotenumformat}%
          \edef\@@foota{\footnotenum=\the\footnotenum\relax}%
          \expandafter\additemR\expandafter\@footup\expandafter
               {\@@foota\footnotenumformat}\to\@footnotelist
          \global\let\@footnotelist=\@footnotelist
     \else\markup{#1}%
          \additemR\markup{#1}\to\@footnotelist
          \global\let\@footnotelist=\@footnotelist
     \fi
     \@footsf}

\def\footnotetext{%
     \ifx\@footnotelist\empty\err@extrafootnotetext\else\@footnotetext\fi}
\def\@footnotetext{%
     \getitemL\@footnotelist\to\@@foota
     \global\let\@footnotelist=\@footnotelist
     \insert\footins\bgroup
     \footnoteformat
     \splittopskip=\ht\strutbox\splitmaxdepth=\dp\strutbox
     \interlinepenalty=\interfootnotelinepenalty\floatingpenalty=\@MM
     \noindent\llap{\@@foota}\strut
     \bgroup\aftergroup\@footnoteend
     \let\@@scratcha=}
\def\@footnoteend{\strut\par\vskip\footnoteskip\egroup}

\def\footnoterule{\normalfonts
     \kern-.3em \hrule width2in height.04em \kern .26em }

\def\footnotecheck{%
     \ifx\@footnotelist\empty
     \else\err@extrafootnotemark
          \global\let\@footnotelist=\empty
     \fi}

\message{labels,}

\let\@@labeldef=\xdef
\newif\if@labelfile
\newwrite\@labelfile
\let\@prelabellist=\empty

\def\label#1#2{\trim#1\to\@@labarg\edef\@@labtext{#2}%
     \edef\@@labname{lab@\@@labarg}%
     \useafter\ifundefined\@@labname\then\else\@yeslab\fi
     \useafter\@@labeldef\@@labname{#2}%
     \ifstreaming
          \expandafter\toks@\expandafter\expandafter\expandafter
               {\csname\@@labname\endcsname}%
          \immediate\write\streamout{\noexpand\label{\@@labarg}{\the\toks@}}%
     \fi}
\def\@yeslab{%
     \useafter\ifundefined{if\@@labname}\then
          \err@labelredef\@@labarg
     \else\useif{if\@@labname}\then
               \err@labelredef\@@labarg
          \else\global\usename{\@@labname true}%
               \useafter\ifundefined{pre\@@labname}\then
               \else\useafter\ifx{pre\@@labname}\@@labtext
                    \else\err@badlabelmatch\@@labarg
                    \fi
               \fi
               \if@labelfile
               \else\global\@labelfiletrue
                    \immediate\write\sixt@@n{--> Creating file \jobname.lab}%
                    \immediate\openout\@labelfile=\jobname.lab
               \fi
               \immediate\write\@labelfile
                    {\noexpand\prelabel{\@@labarg}{\@@labtext}}%
          \fi
     \fi}

\def\putlab#1{\trim#1\to\@@labarg\edef\@@labname{lab@\@@labarg}%
     \useafter\ifundefined\@@labname\then\@nolab\else\usename\@@labname\fi}
\def\@nolab{%
     \useafter\ifundefined{pre\@@labname}\then
          \undefinedlabelformat
          \err@needlabel\@@labarg
          \useafter\xdef\@@labname{\undefinedlabelformat}%
     \else\usename{pre\@@labname}%
          \useafter\xdef\@@labname{\usename{pre\@@labname}}%
     \fi
     \useafter\newif{if\@@labname}%
     \expandafter\additemR\@@labarg\to\@prelabellist}

\def\prelabel#1{\useafter\gdef{prelab@#1}}

\def\ifundefinedlabel#1\then{%
     \expandafter\ifx\csname lab@#1\endcsname\relax}
\def\useiflab#1\then{\csname iflab@#1\endcsname}

\def\prelabelcheck{{%
     \def\^^\##1{\useiflab{##1}\then\else\err@undefinedlabel{##1}\fi}%
     \@prelabellist}}

\message{equation numbering,}

\newcount\chapternum
\newcount\sectionnum
\newcount\subsectionnum
\newcount\equationnum
\newcount\subequationnum
\newcount\figurenum
\newcount\subfigurenum
\newcount\tablenum
\newcount\subtablenum

\newif\if@subeqncount
\newif\if@subfigcount
\newif\if@subtblcount

\def\newchapternum{\newsectionnum=\z@\@resetnum\chapternum}
\def\newsectionnum{\newsubsectionnum=\z@\@resetnum\sectionnum}
\def\newsubsectionnum{\newequationnum=\z@\newfigurenum=\z@\newtablenum=\z@
     \@resetnum\subsectionnum}
\def\newequationnum{\newsubequationnum=\z@\@resetnum\equationnum}
\def\newsubequationnum{\@resetnum\subequationnum}
\def\newfigurenum{\newsubfigurenum=\z@\@resetnum\figurenum}
\def\newsubfigurenum{\@resetnum\subfigurenum}
\def\newtablenum{\newsubtablenum=\z@\@resetnum\tablenum}
\def\newsubtablenum{\@resetnum\subtablenum}

\def\@resetnum#1{\global\advance#1by1 \edef\next{\the#1\relax}\global#1}

\newchapternum=0

\def\chapternumstyle#1{\@setnumstyle\chapternum{#1}}
\def\sectionnumstyle#1{\@setnumstyle\sectionnum{#1}}
\def\subsectionnumstyle#1{\@setnumstyle\subsectionnum{#1}}
\def\equationnumstyle#1{\@setnumstyle\equationnum{#1}}
\def\subequationnumstyle#1{\@setnumstyle\subequationnum{#1}%
     \ifblank\subequationnumstyle\then\global\@subeqncountfalse\fi
     \ignorespaces}
\def\figurenumstyle#1{\@setnumstyle\figurenum{#1}}
\def\subfigurenumstyle#1{\@setnumstyle\subfigurenum{#1}%
     \ifblank\subfigurenumstyle\then\global\@subfigcountfalse\fi
     \ignorespaces}
\def\tablenumstyle#1{\@setnumstyle\tablenum{#1}}
\def\subtablenumstyle#1{\@setnumstyle\subtablenum{#1}%
     \ifblank\subtablenumstyle\then\global\@subtblcountfalse\fi
     \ignorespaces}

\def\eqnlabel#1{%
     \if@subeqncount
          \newsubequationnum=\next
     \else\newequationnum=\next
          \ifblank\subequationnumstyle\then
          \else\global\@subeqncounttrue
               \newsubequationnum=\@ne
          \fi
     \fi
     \label{#1}{\puteqnformat}(\puteqn{#1})%
     \ifdraft\rlap{\hskip.1in{\tt#1}}\fi}

\let\puteqn=\putlab

\def\equation#1#2{\useafter\gdef{eqn@#1}{#2\eqno\eqnlabel{#1}}}
\def\Equation#1{\useafter\gdef{eqn@#1}}

\def\putequation#1{\useafter\ifundefined{eqn@#1}\then
     \err@undefinedeqn{#1}\else\usename{eqn@#1}\fi}

\def\eqnseriesstyle#1{\gdef\@eqnseriesstyle{#1}}
\def\begineqnseries{\subequationnumstyle{\@eqnseriesstyle}%
     \defaultoption[]\@begineqnseries}
\def\@begineqnseries[#1]{\edef\@@eqnname{#1}}
\def\endeqnseries{\subequationnumstyle{blank}%
     \expandafter\ifnoarg\@@eqnname\then
     \else\label\@@eqnname{\puteqnformat}%
     \fi
     \aftergroup\ignorespaces}

\def\figlabel#1{%
     \if@subfigcount
          \newsubfigurenum=\next
     \else\newfigurenum=\next
          \ifblank\subfigurenumstyle\then
          \else\global\@subfigcounttrue
               \newsubfigurenum=\@ne
          \fi
     \fi
     \label{#1}{\putfigformat}\putfig{#1}%
     {\def\marginnoteformat{\tt}\marginnote{#1}}}

\let\putfig=\putlab

\def\figseriesstyle#1{\gdef\@figseriesstyle{#1}}
\def\beginfigseries{\subfigurenumstyle{\@figseriesstyle}%
     \defaultoption[]\@beginfigseries}
\def\@beginfigseries[#1]{\edef\@@figname{#1}}
\def\endfigseries{\subfigurenumstyle{blank}%
     \expandafter\ifnoarg\@@figname\then
     \else\label\@@figname{\putfigformat}%
     \fi
     \aftergroup\ignorespaces}

\def\tbllabel#1{%
     \if@subtblcount
          \newsubtablenum=\next
     \else\newtablenum=\next
          \ifblank\subtablenumstyle\then
          \else\global\@subtblcounttrue
               \newsubtablenum=\@ne
          \fi
     \fi
     \label{#1}{\puttblformat}\puttbl{#1}%
     {\def\marginnoteformat{\tt}\marginnote{#1}}}

\let\puttbl=\putlab

\def\tblseriesstyle#1{\gdef\@tblseriesstyle{#1}}
\def\begintblseries{\subtablenumstyle{\@tblseriesstyle}%
     \defaultoption[]\@begintblseries}
\def\@begintblseries[#1]{\edef\@@tblname{#1}}
\def\endtblseries{\subtablenumstyle{blank}%
     \expandafter\ifnoarg\@@tblname\then
     \else\label\@@tblname{\puttblformat}%
     \fi
     \aftergroup\ignorespaces}

\message{reference numbering,}

\newcount\referencenum \referencenum=0
\newcount\@@prerefcount \@@prerefcount=0
\newcount\@@thisref
\newcount\@@lastref
\newcount\@@loopref
\newcount\@@refseq
\newdimen\refnumindent
\let\@undefreflist=\empty

\def\referencenumstyle#1{\@setnumstyle\referencenum{#1}}

\def\referencestyle#1{\usename{@ref#1}}

\def\@refsequential{%
     \gdef\@refpredef##1{\global\advance\referencenum by\@ne
          \let\^^\=0\label{##1}{\^^\{\the\referencenum}}%
          \useafter\gdef{ref@\the\referencenum}{{##1}{\undefinedlabelformat}}}%
     \gdef\@reference##1##2{%
          \ifundefinedlabel##1\then
          \else\def\^^\####1{\global\@@thisref=####1\relax}\putlab{##1}%
               \useafter\gdef{ref@\the\@@thisref}{{##1}{##2}}%
          \fi}%
     \gdef\endputreferences{%
          \loop\ifnum\@@loopref<\referencenum
                    \advance\@@loopref by\@ne
                    \expandafter\expandafter\expandafter\@printreference
                         \csname ref@\the\@@loopref\endcsname
          \repeat
          \par}}

\def\@refpreordered{%
     \gdef\@refpredef##1{\global\advance\referencenum by\@ne
          \additemR##1\to\@undefreflist}%
     \gdef\@reference##1##2{%
          \ifundefinedlabel##1\then
          \else\global\advance\@@loopref by\@ne
               {\let\^^\=0\label{##1}{\^^\{\the\@@loopref}}}%
               \@printreference{##1}{##2}%
          \fi}
     \gdef\endputreferences{%
          \def\^^\####1{\useiflab{####1}\then
               \else\reference{####1}{\undefinedlabelformat}\fi}%
          \@undefreflist
          \par}}

\def\beginprereferences{\par
     \def\reference##1##2{\global\advance\referencenum by1\@ne
          \let\^^\=0\label{##1}{\^^\{\the\referencenum}}%
          \useafter\gdef{ref@\the\referencenum}{{##1}{##2}}}}
\def\endprereferences{\global\@@prerefcount=\the\referencenum\par}

\def\beginputreferences{\par
     \refnumindent=\z@\@@loopref=\z@
     \loop\ifnum\@@loopref<\referencenum
               \advance\@@loopref by\@ne
               \setbox\z@=\hbox{\referencenum=\@@loopref
                    \referencenumformat\enskip}%
               \ifdim\wd\z@>\refnumindent\refnumindent=\wd\z@\fi
     \repeat
     \putreferenceformat
     \@@loopref=\z@
     \loop\ifnum\@@loopref<\@@prerefcount
               \advance\@@loopref by\@ne
               \expandafter\expandafter\expandafter\@printreference
                    \csname ref@\the\@@loopref\endcsname
     \repeat
     \let\reference=\@reference}

\def\@printreference#1#2{\ifx#2\undefinedlabelformat\err@undefinedref{#1}\fi
     \noindent\ifdraft\rlap{\hskip\hsize\hskip.1in \tt#1}\fi
     \llap{\referencenum=\@@loopref\referencenumformat\enskip}#2\par}

\def\reference#1#2{{\par\refnumindent=\z@\putreferenceformat\noindent#2\par}}

\def\putref#1{\trim#1\to\@@refarg
     \expandafter\ifnoarg\@@refarg\then
          \toks@={\relax}%
     \else\@@lastref=-\@m\def\@@refsep{}\def\@more{\@nextref}%
          \toks@={\@nextref#1,,}%
     \fi\the\toks@}
\def\@nextref#1,{\trim#1\to\@@refarg
     \expandafter\ifnoarg\@@refarg\then
          \let\@more=\relax
     \else\ifundefinedlabel\@@refarg\then
               \expandafter\@refpredef\expandafter{\@@refarg}%
          \fi
          \def\^^\##1{\global\@@thisref=##1\relax}%
          \global\@@thisref=\m@ne
          \setbox\z@=\hbox{\putlab\@@refarg}%
     \fi
     \advance\@@lastref by\@ne
     \ifnum\@@lastref=\@@thisref\advance\@@refseq by\@ne\else\@@refseq=\@ne\fi
     \ifnum\@@lastref<\z@
     \else\ifnum\@@refseq<\thr@@
               \@@refsep\def\@@refsep{,}%
               \ifnum\@@lastref>\z@
                    \advance\@@lastref by\m@ne
                    {\referencenum=\@@lastref\putrefformat}%
               \else\undefinedlabelformat
               \fi
          \else\def\@@refsep{--}%
          \fi
     \fi
     \@@lastref=\@@thisref
     \@more}

\message{streaming,}

\newif\ifstreaming

\def\streamto{\defaultoption[\jobname]\@streamto}
\def\@streamto[#1]{\global\streamingtrue
     \immediate\write\sixt@@n{--> Streaming to #1.str}%
     \newwrite\streamout\immediate\openout\streamout=#1.str }

\def\streamfrom{\defaultoption[\jobname]\@streamfrom}
\def\@streamfrom[#1]{\newread\streamin\openin\streamin=#1.str
     \ifeof\streamin
          \expandafter\err@nostream\expandafter{#1.str}%
     \else\immediate\write\sixt@@n{--> Streaming from #1.str}%
          \let\@@labeldef=\gdef
          \ifstreaming
               \edef\@elc{\endlinechar=\the\endlinechar}%
               \endlinechar=\m@ne
               \loop\read\streamin to\@@scratcha
                    \ifeof\streamin
                         \streamingfalse
                    \else\toks@=\expandafter{\@@scratcha}%
                         \immediate\write\streamout{\the\toks@}%
                    \fi
                    \ifstreaming
               \repeat
               \@elc
               \input #1.str
               \streamingtrue
          \else\input #1.str
          \fi
          \let\@@labeldef=\xdef
     \fi}

\def\streamcheck{\ifstreaming
     \immediate\write\streamout{\pagenum=\the\pagenum}%
     \immediate\write\streamout{\footnotenum=\the\footnotenum}%
     \immediate\write\streamout{\referencenum=\the\referencenum}%
     \immediate\write\streamout{\chapternum=\the\chapternum}%
     \immediate\write\streamout{\sectionnum=\the\sectionnum}%
     \immediate\write\streamout{\subsectionnum=\the\subsectionnum}%
     \immediate\write\streamout{\equationnum=\the\equationnum}%
     \immediate\write\streamout{\subequationnum=\the\subequationnum}%
     \immediate\write\streamout{\figurenum=\the\figurenum}%
     \immediate\write\streamout{\subfigurenum=\the\subfigurenum}%
     \immediate\write\streamout{\tablenum=\the\tablenum}%
     \immediate\write\streamout{\subtablenum=\the\subtablenum}%
     \immediate\closeout\streamout
     \fi}


\def\err@badtypesize{%
     \errhelp={The limited availability of certain fonts requires^^J%
          that the base type size be 10pt, 12pt, or 14pt.^^J}%
     \errmessage{--> Illegal base type size}}

\def\err@badsizechange{\immediate\write\sixt@@n
     {--> Size change not allowed in math mode, ignored}}

\def\err@sizetoolarge#1{\immediate\write\sixt@@n
     {--> \noexpand#1 too big, substituting HUGE}}

\def\err@sizenotavailable#1{\immediate\write\sixt@@n
     {--> Size not available, \noexpand#1 ignored}}

\def\err@fontnotavailable#1{\immediate\write\sixt@@n
     {--> Font not available, \noexpand#1 ignored}}

\def\err@sltoit{\immediate\write\sixt@@n
     {--> Style \noexpand\sl not available, substituting \noexpand\it}%
     \it}

\def\err@bfstobf{\immediate\write\sixt@@n
     {--> Style \noexpand\bfs not available, substituting \noexpand\bf}%
     \bf}

\def\err@badgroup#1#2{%
     \errhelp={The block you have just tried to close was not the one^^J%
          most recently opened.^^J}%
     \errmessage{--> \noexpand\end{#1} doesn't match \noexpand\begin{#2}}}

\def\err@badcountervalue#1{\immediate\write\sixt@@n
     {--> Counter (#1) out of bounds}}

\def\err@extrafootnotemark{\immediate\write\sixt@@n
     {--> \noexpand\footnotemark command
          has no corresponding \noexpand\footnotetext}}

\def\err@extrafootnotetext{%
     \errhelp{You have given a \noexpand\footnotetext command without first
          specifying^^Ja \noexpand\footnotemark.^^J}%
     \errmessage{--> \noexpand\footnotetext command has no corresponding
          \noexpand\footnotemark}}

\def\err@labelredef#1{\immediate\write\sixt@@n
     {--> Label "#1" redefined}}

\def\err@badlabelmatch#1{\immediate\write\sixt@@n
     {--> Definition of label "#1" doesn't match value in \jobname.lab}}

\def\err@needlabel#1{\immediate\write\sixt@@n
     {--> Label "#1" cited before its definition}}

\def\err@undefinedlabel#1{\immediate\write\sixt@@n
     {--> Label "#1" cited but never defined}}

\def\err@undefinedeqn#1{\immediate\write\sixt@@n
     {--> Equation "#1" not defined}}

\def\err@undefinedref#1{\immediate\write\sixt@@n
     {--> Reference "#1" not defined}}

\def\err@nostream#1{%
     \errhelp={You have tried to input a stream file that doesn't exist.^^J}%
     \errmessage{--> Stream file #1 not found}}

\message{jyTeX initialization}

\everyjob{\immediate\write16{--> jyTeX version \fmtversion}%
     \edef\@@jobname{\jobname}%
     \edef\jobname{\@@jobname}%
     \settime
     \openin0=\jobname.lab
     \ifeof0
     \else\closein0
          \immediate\write16{--> Getting labels from file \jobname.lab}%
          \input\jobname.lab
     \fi}


\def\fixedskipslist{%
     \^^\{\topskip}%
     \^^\{\splittopskip}%
     \^^\{\maxdepth}%
     \^^\{\skip\topins}%
     \^^\{\skip\footins}%
     \^^\{\headskip}%
     \^^\{\footskip}}

\def\scalingskipslist{%
     \^^\{\p@renwd}%
     \^^\{\delimitershortfall}%
     \^^\{\nulldelimiterspace}%
     \^^\{\scriptspace}%
     \^^\{\jot}%
     \^^\{\normalbaselineskip}%
     \^^\{\normallineskip}%
     \^^\{\normallineskiplimit}%
     \^^\{\baselineskip}%
     \^^\{\lineskip}%
     \^^\{\lineskiplimit}%
     \^^\{\bigskipamount}%
     \^^\{\medskipamount}%
     \^^\{\smallskipamount}%
     \^^\{\parskip}%
     \^^\{\parindent}%
     \^^\{\abovedisplayskip}%
     \^^\{\belowdisplayskip}%
     \^^\{\abovedisplayshortskip}%
     \^^\{\belowdisplayshortskip}%
     \^^\{\abovechapterskip}%
     \^^\{\belowchapterskip}%
     \^^\{\abovesectionskip}%
     \^^\{\belowsectionskip}%
     \^^\{\abovesubsectionskip}%
     \^^\{\belowsubsectionskip}}


\def\twoupsetup{
     \topmargin=.75in
     \leftmargin=.5in
     \vsize=6.9in
     \hsize=4.75in
     \fullhsize=10in
     \let\draft=\relax}

\outputstyle{normal}                             

\def\marginnoteformat{\subscriptsize             
     \hsize=1in \baselinestretch=1000 \everypar={}%
     \tolerance=5000 \hbadness=5000 \parskip=0pt \parindent=0pt
     \leftskip=0pt \rightskip=0pt \raggedright}

\head={\ifdraft\normalfonts\it\hfil DRAFT\hfil   
     \llap{\number\day\ \monthword\month\ \militarytime}\else\hfil\fi}
\foot={\hfil\normalfonts\numstyle\pagenum\hfil}  

\normalbaselineskip=12pt                         
\normallineskip=0pt                              
\normallineskiplimit=0pt                         
\normalbaselines                                 

\topskip=.85\baselineskip
\splittopskip=\topskip
\headskip=2\baselineskip
\footskip=\headskip

\pagenumstyle{arabic}                            

\parskip=0pt                                     
\parindent=20pt                                  

\baselinestretch=1000                            


\chapterstyle{left}                              
\chapternumstyle{blank}                          
\def\chapterbreak{\newpage}                      
\abovechapterskip=0pt                            
\belowchapterskip=1.5\baselineskip               
     plus.38\baselineskip minus.38\baselineskip
\def\chapternumformat{\numstyle\chapternum.}     

\sectionstyle{left}                              
\sectionnumstyle{blank}                          
\def\sectionbreak{\vskip0pt plus4\baselineskip\penalty-100
     \vskip0pt plus-4\baselineskip}              
\abovesectionskip=1.5\baselineskip               
     plus.38\baselineskip minus.38\baselineskip
\belowsectionskip=\the\baselineskip              
     plus.25\baselineskip minus.25\baselineskip
\def\sectionnumformat{
     \ifblank\chapternumstyle\then\else\numstyle\chapternum.\fi
     \numstyle\sectionnum.}

\subsectionstyle{left}                           
\subsectionnumstyle{blank}                       
\def\subsectionbreak{\vskip0pt plus4\baselineskip\penalty-100
     \vskip0pt plus-4\baselineskip}              
\abovesubsectionskip=\the\baselineskip           
     plus.25\baselineskip minus.25\baselineskip
\belowsubsectionskip=.75\baselineskip            
     plus.19\baselineskip minus.19\baselineskip
\def\subsectionnumformat{
     \ifblank\chapternumstyle\then\else\numstyle\chapternum.\fi
     \ifblank\sectionnumstyle\then\else\numstyle\sectionnum.\fi
     \numstyle\subsectionnum.}


\footnotenumstyle{symbols}                       
\footnoteskip=0pt                                
\def\footnotenumformat{\numstyle\footnotenum}    
\def\footnoteformat{\footnotesize                
     \everypar={}\parskip=0pt \parfillskip=0pt plus1fil
     \leftskip=1em \rightskip=0pt
     \spaceskip=0pt \xspaceskip=0pt
     \def\\{\ifhmode\ifnum\lastpenalty=-10000
          \else\hfil\penalty-10000 \fi\fi\ignorespaces}}


\def\undefinedlabelformat{$\bullet$}             


\equationnumstyle{arabic}                        
\subequationnumstyle{blank}                      
\figurenumstyle{arabic}                          
\subfigurenumstyle{blank}                        
\tablenumstyle{arabic}                           
\subtablenumstyle{blank}                         

\eqnseriesstyle{alphabetic}                      
\figseriesstyle{alphabetic}                      
\tblseriesstyle{alphabetic}                      

\def\puteqnformat{\hbox{
     \ifblank\chapternumstyle\then\else\numstyle\chapternum.\fi
     \ifblank\sectionnumstyle\then\else\numstyle\sectionnum.\fi
     \ifblank\subsectionnumstyle\then\else\numstyle\subsectionnum.\fi
     \numstyle\equationnum
     \numstyle\subequationnum}}
\def\putfigformat{\hbox{
     \ifblank\chapternumstyle\then\else\numstyle\chapternum.\fi
     \ifblank\sectionnumstyle\then\else\numstyle\sectionnum.\fi
     \ifblank\subsectionnumstyle\then\else\numstyle\subsectionnum.\fi
     \numstyle\figurenum
     \numstyle\subfigurenum}}
\def\puttblformat{\hbox{
     \ifblank\chapternumstyle\then\else\numstyle\chapternum.\fi
     \ifblank\sectionnumstyle\then\else\numstyle\sectionnum.\fi
     \ifblank\subsectionnumstyle\then\else\numstyle\subsectionnum.\fi
     \numstyle\tablenum
     \numstyle\subtablenum}}


\referencestyle{sequential}                      
\referencenumstyle{arabic}                       
\def\putrefformat{\numstyle\referencenum}        
\def\referencenumformat{\numstyle\referencenum.} 
\def\putreferenceformat{
     \everypar={\hangindent=1em \hangafter=1 }%
     \def\\{\hfil\break\null\hskip-1em \ignorespaces}%
     \leftskip=\refnumindent\parindent=0pt \interlinepenalty=1000 }


\normalsize


\def\fmtversion{2.6M (June 1992)}

\catcode`\@=12

%
%
\def\h{\hbox to .5cm{\hfill}}
\def\hof{\hbox to .15cm{\hfill}}
\def\hi{\hbox to .2cm{\hfill}}
\def\htf{\hbox to .35cm{\hfill}}

\def\ha#1{\n\hbox to .6cm{\n {#1}\hfill}}
\def\hb#1{\indent\hbox to .7cm{\n {#1}\hfill}}
\def\hc#1{\indent\hbox to .7cm{\hfill}\hbox to 1.1cm{\n {#1}\hfill}}
\def\hd#1{\indent\hbox to 1.8cm{\hfill}\hbox to 1.4cm{\n {#1}\hfill}}

\def\eq#1{\eqno\eqnlabel{#1}}

\def\mpr#1{\markup{[\putref{#1}]}}
\def\pr#1{[\putref{#1}]}

\def\n{\noindent}
\def\no{\noindent}
\def\ref{\reference}
\def\referencenumformat{[\numstyle\referencenum]}

\def\undbib{\underbar {\hbox to 2cm{\hfill}}, }
%
%

\def\dag{ \dagger }
\def\dc{\tilde h}

\def\e{{\rm e}}

\def\lal{{\cal L}}

\def\theta{\vartheta}

%
%
\def\half{{1\over 2}}

%
%

\def\ie{{\it i.e.}}

%
%

\begin{ignore}

\font\twelveBbb=msym10 scaled \magstep1
\font\nineBbb=msym9
\font\sevenBbb=msym7
\newfam\Bbbfam
\textfont\Bbbfam=\twelveBbb
\scriptfont\Bbbfam=\nineBbb
\scriptscriptfont\Bbbfam=\sevenBbb
\def\Bbb{\fam\Bbbfam\twelveBbb}

  \def\S{{\Bbb S}} 
   
 \def\Z{{\Bbb Z}}
\end{ignore}
\def\Z{Z\!\!\!Z}


\topmargin=1truein\vsize=9truein
\leftmargin=1truein\hsize=6.5truein

\baselinestretch=1200

\foot={\hfil\normalfonts\numstyle\pagenum\hfil}
\head={\hfil}

\pagenum=0
\pagenumstyle{arabic}
\footnotenumstyle{arabic}
\footnotenum= 0

\sectionnumstyle{arabic}
\subsectionnumstyle{blank}
\sectionnum=0

\pagenumstyle{blank}

\pagenum=0\pagenumstyle{blank}

{\pagenumstyle{blank}

{\rightline{\hbox to 4.5cm{\vtop{\hsize= 4.5cm
\baselinestretch=960\footnotefonts
\hfill\\
OHSTPY-HEP-T-94-007\\
DOE/ER/01545-626\\
hep-th/9409096\\
August 1994
}}{\hbox to .35cm{\hfill}}}}

\vskip 1.5 truecm
\footnotenum= 0

\centertext{{\Bigfonts\bf GUT's With Adjoint Higgs Fields\\From Superstrings}}

\vskip 1.2 truecm

\centertext{{Gerald B.~Cleaver}}
\vskip .3 truecm
\centertext{{\it Department of Physics, The Ohio State University,\\
174 W.~18th Ave., Columbus, OH 43210\\
gcleaver@pacific.mps.ohio-state.edu}}

\vskip 2.0 truecm
\centerline{\bf Abstract}
\vskip .3 truecm
{\parindent=3pc
\typesize=10pt
\begin{narrow}
I discuss how effective grand unified theories
requiring adjoint Higgs fields
for breaking to the standard model
can be contained within string theory.
Initial findings are presented in a search
for and classification of effective
three generation SO(10) SUSY-GUT models built using
the free-fermionic string approach.
\end{narrow}}
\hbox to 1truecm{\hfill}
\centertext{\it Based on talk presented at PASCOS '94, Syracuse, NY}
\hfill\vfill\eject}

\pagenum=0\pagenumstyle{arabic}
\footnotenumstyle{symbols}

\chapternumstyle{blank}
\subsectionnumstyle{blank}

\no{\bf 1. }{\bf GUTs and Strings}\vskip .4cm
\sectionnumstyle{arabic}
\sectionnum=1\equationnum=0

Elementary particle physics has achieved phenomenal success in recent
decades, resulting in the Standard Model (SM),
SU(3)$_{\rm C}\times$SU(2)$_{\rm L}\times$U(1)$_{\rm Y}$,
and verification to high
precision of its many predictions. However, there are still several
unsatisfying aspects of the theory:
(1) The SM is very complicated, requiring measurement of some 19 free
parameters, such as the masses of the quarks and leptons
and the coupling constants.  We should expect the true fundamental theory to
have at most one free parameter.
(2) A gauge group
that is the direct product of three gauge groups with independent couplings
does not seem fundamental.
(3) There is a naturalness problem concerning
 the scale at which the
electroweak (EW) symmetry, SU(2)$_{\rm L}\times$U(1)$_{\rm Y}$, breaks to the
electromagnetic U(1)$_{\rm EM}$.
Although this is ``explained'' by the scale of the
Higgs mass, fine-tuning is required in renormalization theory to keep
the Higgs mass on the order of the symmetry breaking scale, which
suggests the need for supersymmetry at a higher scale.
(4) Fine-tuning is also required to solve the strong CP problem.
(5) The SM provides no unification with gravity, {\ie}, no means of
forming a consistent theory of quantum gravity.
(6) The cosmological constant resulting from EW symmetry breaking
should be many orders of magnitude higher than the experimental
limit, which again necessitates fine-tuning cancellation.

These shortcomings have
motivated a search for phenomenologically
viable
Grand Unified Theories (GUTs) that would unify
SM physics through a single force and/or even for a Theory of Everything
(TOE) that
could consistently combine the SM with gravity.  In the last decade,
this pursuit has resulted in an intensive study of string theory.
String theory is the first theory to successfully
combine the SM forces with gravity.

In one sense, string theory has been too successful following
the explosion of interest in the mid-80's.  The (super)string
theory is inherently a (10) 26 dimensional spacetime theory. Although
only a few perturbative solutions to the theories exist when all
spacetime dimensions are uncompactified, for every compactified dimension
there arises many more possible solutions.
With only four uncompactified
spacetime dimensions, there is a plethora
of distinct solutions to the superstring theory.
Many different approaches to ``compactification,'' {\it e.g.},
bosonic lattices and orbifolds, free fermions, Calabi-Yau manifolds,
and $N=2$ minimal models, have been devised.
(Often there is much overlap and
sometimes even complete equivalence between varying methods of
compactification.)  Four-dimensional solutions
can be classified into two broad categories: (1) those
involving an actual geometrical
compactification from ten uncompactified dimensions,
and (2) those with
internal degrees of freedom having no equivalent representation
in terms of six well-defined compactified dimensions.

There is a
potential problem with solutions in the first class: such models with
$N=1$ spacetime supersymmetry (SUSY) and/or chiral fermions cannot
contain massless spacetime scalar fields in adjoint or higher dimensional
representations of the gauge group.\mpr{lewellen,font90,ellis90}
This has presented a possible difficulty for string theory, because typical
GUTs depend  upon scalars in these representations to break the
gauge symmetry down to the  SM. In the usual approach,
spontaneous symmetry breaking is brought about
by vacuum expectation values (VEVs) of these scalars.
Therefore, either the gauge groups of these string models must break to
the standard model near the string (Planck) scale or a non-standard
Higgs breaking is required.  An example of the first method is symmetry
breaking by Wilson lines in Calabi-Yau vacua.\mpr{candelas85}
Flipped $SU(5)$ is the primary example of the second
approach.\mpr{antoniadis87b}
However, standard GUTs such as SU(5) or SO(10)
are excluded from this class of string theory models.

In the first class of models,
the absence of spacetime scalars in higher representations results
from the association of geometrical compactification with level-one
Ka\v c-Moody algebras (KMAs).
Because of this connection,
basing a model on level-one KMAs has been
the standard approach to string theory phenomenology.
Starting from
either the ten dimensional type-II or heterotic superstrings,
four-dimensional spacetime has most often been derived through
``spontaneous compactification'' of the extra six dimensions.
In ten uncompactified
dimensions the only modular invariant heterotic
string models with spacetime SUSY and gauge symmetry are the
level-one $E_8\otimes E_8$ and level-one SO(32) solutions.
(In ten uncompactified dimensions, the type-II string has $N=2$ SUSY,
but no gauge group.)
Compactification of the extra six dimensions
on a Calabi-Yau manifold or symmetric orbifold,
naturally keeps
the KMA at level-one. The resulting gauge group
is a subgroup of either E$_8\otimes$E$_8$ or SO(32)
(with additional U(1) factors).
Models using bosonic lattice compactification, or equivalently
complex world sheet fermions,\mpr{kawai87a,antoniadis87,antoniadis88}
likewise have level-one
KMAs, with the associated gauge group being a subgroup of
either SO(12)$\otimes$E$_8\otimes$E$_8$ or SO(44).

Models can be based on higher-level KMAs, when the demand for
a classical interpretation of compactification is relaxed.
Such models fall into the second general
class of string solutions and can contain scalars in adjoint or higher
representations.
These states can exist in the spectrum if
their gauge group arises from a level-$K\geq 2$ KMA on
the world sheet.
Model examples using are given in \pr{lewellen}.

When we consider the whole ocean of models containing both
level-one and higher level KMAs, it is perhaps fortunate that
recent LEP results tighten constraints for viable
string models.
Using renormalization group equations (RGEs), the measured high precision
values of
the standard model coupling constants have been extrapolated
from $M_Z$ to near the Planck scale.  It was found that the RGE for the
minimal supersymmetric standard model with just two Higgs doublets predict
a unification of the three coupling constants $g_3$, $g_2$ and $g_1$ for
$SU(3)\times SU(2)_L\times U(1)_Y$, respectively,
at about $10^{16}$ GeV.\mpr{langacker92}  For string theory
this naively poses a problem since the string unification scale is generally
required, at tree level,  to be near the Planck scale
(around $10^{18}$ GeV).
Three classes of solutions have been proposed for resolving
the potential inconsistency between these  extrapolations and string
theory.\markup{[\putref{bailin92}]}
The first proposal is to regard the unification of the couplings at
$10^{16}$ GeV using the minimal SUSY standard model RGE as a coincidence,
 and to allow for additional states  between the electroweak scale and
the string unification scale that raise the RGE unification scale.
A second suggestion is that string threshold effects could
significantly lower the
string scale down to the  minimal SUSY standard model RGE unification
scale.
The third possibility is that a grand unified gauge group
(such as SU(5), SO(10), or E$_6$)
results from a KMA at level $K\geq 2$.\footnote{From hereon, the level
of an SU(N) or SO(2N) KMA is denoted by a subscript, {\it e.g.,}
SO(10)$_2$ for level-2.}
Thus, the SUSY standard model couplings could
unify around $10^{16}$ GeV and run upward from there
with a common value to the string unification scale.

The last proposal appears most natural and
appealing. A grand unified gauge group
fits well with the concept of successive levels of increasing
symmetry much better than does
going directly from the symmetry of the standard model to the symmetry of
the string.  It seems far more natural for the strong force to merge
with the electroweak significantly below the string scale,
rather than at the string scale where the gravitational and hidden sector
gauge couplings finally merge also.

Embedding SUSY GUTs in a string model has several other advantages.
First, the ratio of the SUSY GUT scale to the string scale,
$M_{\rm GUT}/M_{\rm string} \sim 1/100$,
naturally explains the generation hierarchy when only the
heaviest generation obtains mass from dimension four operators derived from
the effective GUT superpotential.
At the GUT scale, (4+n)-dimensional mas operators resulting from
non-perturbative
terms in the superpotential include ratios of VEVs proportional to
$(M_{\rm GUT}/M_{\rm string})^n$. Therefore, from this approach
the first and second
generation mass scales, in comparison to the third,
 can result from five- and six-dimensional operators.

In recent work by Anderson {\it et. al}\mpr{raby}
 the minimum mass texture
of an SO(10) SUSY-GUT needed to reproduce SM physics at the low
energy scale
was determined. Only a few higher (five- and six-) dimensional
operators were needed, far fewer operators than allowed simply from
SUSY-SO(10) symmetry considerations. The additional
symmetries needed to explain the low number of operators in such
models comes naturally when the GUT is embedded in a string theory.
Four-dimensional string models are famous for possessing
extra U(1) factors (as the model in section {\bf 4} demonstrates),
often associated with generation number, and
additional (discrete) worldsheet symmetries that will significantly
reduce the number of permitted terms in the effect GUT superpotential.

As noted,
construction of string models with higher-level gauge groups requires
asymmetry between the left- and right-moving (LM and RM) fields on the
world-sheet.\mpr{lewellen}
Associated with this property of the fields
are asymmetric modular invariants.
Systematically constructing asymmetric modular invariants (AMIs) has proven
very difficult, except for the special case of models based
on free bosons or fermions.\mpr{clgan}
However, even for asymmetric models,
use of lattice bosons (or equivalently complex fermions) limits
the possibilities to level one-models.
The first and simplest alternative
involves using real fermions
(meaning periodic or antiperiodic fermions
with conformal dimensions $\half$) that
cannot be paired with any other real fermions in the model to form a complex
periodic or antiperiodic fermion of complex dimension 1.\mpr{kawai87b}
Such unpairable real fermions assume the role of increasing the
central charge of a group of fermions without increasing the
number of local U(1) charges on the fermions. This parallels the
effect of increasing the level, $K$, of a KM algebra.
Using free fermions is not the
only well-understood method for constructing higher level string models.
As demonstrated in ref.~\pr{font90}
increasing the level of an algebra can also be
brought about through orbifolding, wherein the direct product of
$n$ factors of a gauge group $\mit G$ is modded by a diagonal
$\Z_n$ symmetry. In some cases there is overlap between the two
methods, {\it e.g.}, the SO(10)$_2\otimes$U(1) KMA in the free fermionic
models discussed below can equivalently be produced by orbifolding a
SO(10)$_1\otimes$SO(10)$_1$ manifold by a diagonal $\Z_2$ symmetry.

Actual construction of a free fermionic GUT model
containing adjoint Higgs based on a SO(10)$_2$ KMA
was presented in
ref.~\pr{lewellen}. As this research revealed, the difficulty
with such models is not getting adjoint scalars, but rather
is getting exactly three chiral generations of SM fermions
when the model contains adjoint scalars.
This is not impossible though, as I show in section
{\bf 4} by counter-example.
However, before demonstrating this,
I review in section {\bf 2} the general
constraints on consistent higher-level string models constructed
using any method, along with the specific constraints arising in
the free fermionic approach. Then in section {\bf 3} I review the
fundamentals of free fermionic models.


\vskip .3cm
\no{\bf 2. }{\bf Constraints on Higher-Level GUT Models}
\vskip .2cm
\sectionnum=2\equationnum=0

The first major constraint on string models containing
a level-$K$ KMA comes from the unitarity demand.
This requires that
the representations of the KMA appearing in the string model
satisfy the rule that
$$ K\geq \sum_{i=1}^{{\rm rank}\, \lal}
n_i m_i\,\, , \eqno\eqnlabel{unitaryrep}$$
where $n_i$ are the Dynkin labels of the highest weight representation of the
associated Lie algebra, $\lal$,and $m_i$ are the related co-marks.
Therefore, at level-one only the singlet, spinor,
conjugate spinor and vector representations of SO(4N+2) can
appear, {\it e.g.}, for SO(10)$_1$ only $1$, $10$, $16$, and
$\overline{16}$ representations can appear. Likewise
for SU(N) the only permitted states are
the ${N\choose 0}$- ({\it i.e.,} the singlet),
${N\choose 1}$-, ${N\choose 2}$-, $\dots$,
${N\choose N-1}$-dimensional representations;
and, similarly, for E$_6$ level-one there are only the
$1$, $27$ and $\overline{27}$.
Thus adjoint Higgs appear only when $K\geq 2$.
Naively, there would appear a way of
escaping this. Since the KM currents transform in the adjoint
representation we might use them to form spacetime scalars.
Unfortunately, this is forbidden by the presence of
{\it chiral} fermions and/or $N=1$ spacetime
SUSY.\mpr{dixon87,dreiner89b,lewellen}

The central charge $c_{KMA}({\rm level}-K)$ of an individual
level-$K$ KMA
measures the KMAs contribution to the conformal anomaly of the world
sheet theory. Since the gauge groups originate in the
bosonic sector of the heterotic string, the total contribution to the
conformal anomaly from the gauge groups cannot exceed 22,
$$ \sum_i c_{KMA_i}({\rm level}-K_i) =
\sum_i { K_i{\rm dim}\, \lal_i\over K_i + \dc_i}\leq 22\,\, ,
\eq{centralcharge}$$
where the sum is over the different factors in the algebra
and every U(1)$_K$ contributes 1 to the sum.
$\tilde h_i$ is the dual Coxeter number of the simple Lie Algebra $\lal_i$
embedded in the KMA.
(For simply-laced groups $\tilde h$ equals dim$\,\lal$/rank$\,\lal - 1$,
which for level-one results in a central charge equal to the rank of the
group at level-one.)
This constraint places upper bounds
of 55, seven, and four on the
permitted levels for SU(5), SO(10), and E$_6$ string-based GUTs,
respectively:\mpr{font90,ellis90}
Using
the free fermionic approach places a further constraint on
the allowed levels. Since real, free fermions have central charge
${\half}$, the central charge of the KMA must be an integer multiple of
${\half}$. This restricts the allowed levels to be in the sets
$\{1,3,5,7,10,11,15,19,25,35,43,55\}$, $\{1,2,4,7\}$, and
$\{1,4\}$ for SU(5), SO(10), E$_6$, respectively,
unless there is a compensating contribution to the central charge from
the hidden sector KMA and/or non-KMA factors.

There is one additional constraint:
since the intercept for the bosonic sector of a heterotic string is one,
a potentially massless state in a  representation $(r)$ of the gauge
group cannot have
a conformal dimension $h_{(r)}$ greater than one. That is,
$$h_{(r)}= {C_{(r)}/{\bmit \psi}^2\over K + \dc}\leq 1\,\, ,\eq{maxcdone}$$
where
$C_{(r)}$ is the quadratic Casimir of the representation $(r)$, and
${\bmit \psi}^2$ is the length-squared of the longest root of $\lal$.
Thus, {\it e.g.}, for SO(10)$_2$ no representation above a 54 can be massless,
although representations up to 210 could appear in a unitary level-2 model. Not
until
level-5 can the 126 be massless, but then many unwanted
exotics will be massless as well.
By comparing the needed massless states
to the undesirable massless states potentially
present at a given level,
I believe level-2 is the best choice among the four
potential levels of SO(10) free fermionic models.
Note that massless 126 scalars need not be present in
SO(10) models for good phenomenology. The highest scalar needed is
actually only the 54.\mpr{raby}

\vskip .3cm
\no{\bf 3. }{\bf Free Fermionic Models}
\vskip .2cm
\sectionnum=3\equationnum=0

Free fermionic model building was developed simultaneously by
Kawai, Lewellen, and Tye in
\pr{kawai87a} and by Antoniadis, Bachas, and Kounas in \pr{antoniadis87}
and further advanced by these two groups in \pr{kawai87b} and
\pr{antoniadis88}.
In light-cone gauge, a free fermionic heterotic string model contains
64 real worldsheet fermions $\psi_n$
($1\leq n \leq 20$ for LM, $21\leq n \leq 64$ RM)
in addition to the LM and RM worldsheet scalars
($X^{i}\, ,\bar X^{\bar j}$) embedding transverse coordinates of
four-dimensional spacetime. $\psi_1$ and $\psi_2$ are the worldsheet
superpartners of the two LM transverse scalars; the
remaining 62 are internal degrees of freedom. A specific model is
defined by (1) sets of 64-component boundary vectors
describing how the
fermions transform around non-contractible loops on the worldsheet,
and (2) sets of coefficients weighting contributions to a partition
function from fermions with specific boundary conditions.

Modular invariance is a requirement for a sensible model and exists
if (1) the one-loop partition function is invariant under $S$ and $T$
transformations of the complex worldsheet parameter $\tau$ and (2) either
a specific additional two-loop constraint is
satisfied\mpr{antoniadis88}
or, equivalently,
the states surviving the one-loop GSO projection ``are
sensible''\mpr{kawai87a}.
The one-loop worldsheet is described by a torus and, therefore, provides two
non-contractible loops around which a fermion may be transported.
The transformation properties for any one of the 64 real
fermions $\psi_{\rm n}$
after going around either non-contractible
loops may be expressed as
$$\psi_{n} \rightarrow - \exp\{\pi \, i\,\alpha_n\}\psi_n\,\,\,
{\rm ~or~}\,\,\,
  \psi_{n} \rightarrow - \exp\{\pi \, i\,\beta_n\}\psi_n\,\, ,
\eq{trans1-b}$$
respectively, where $-1<\alpha_n,\, \beta_n\leq 1$.
$\alpha_n$ and $\beta_n$ are the n$^{\rm th}$ components
of 64-dimensional vectors ${\bmit \alpha}$ and ${\bmit \beta}$,
respectively. A real fermion $\psi_n$ may have only periodic
(Ramond fermion) or
antiperiodic boundary conditions (Neveu-Schwarz fermion)
around each loop, {\it i.e.},
$\alpha_n,\, \beta_n \in \{ 1  {\rm ~(periodic)~},\, 0 {\rm~(antiperiodic)}\}$.
Alternatively, a LM (RM) $\psi_n$ may be paired with another
LM (RM) real fermion $\psi_m$
to form a Weyl fermion $\psi_{n,m}\equiv \psi_n + i\psi_m$
with complex boundary conditions around the loops.
This is allowed when both fermions have identical periodic/antiperiodic
boundary conditions everywhere; then
$\alpha_n=\alpha_m\equiv\alpha_{n,m},\,
\beta_n=\beta_m\equiv\beta_{n,m}$ can be rational in sectors.

The contribution to the one-loop partition function, $Z_{\rm fermion}$,
from the 64 real fermions with their chosen sets of boundary vectors,
$\{\alpha\}$ and $\{\beta\}$, can be expressed as a weighted summation
over the individual partition functions for a specific pair of
boundary vectors,
$$Z_{\rm fermion}= \sum_{{\bmit\alpha}\in\{ {\bmit\alpha}\}\atop
                         {\bmit\beta}\in\{ {\bmit\beta}\} }
C\left({\bmit\alpha\atop\bmit\beta}\right)
Z\left({\bmit\alpha\atop\bmit\beta}\right)\,\, .
\eq{partfunc}$$
The weights $C\left({\bmit\alpha\atop\bmit\beta}\right)$ are complex
phases if either $\alpha$ or $\beta$ have rational, non-integer components
and are real phases ($\pm 1$) when both $\alpha$ and $\beta$
are integer. One-loop modular invariance requires that
$\{ {\bmit\alpha}\}$ and $\{ {\bmit\beta}\}$ be identical sets and that if
$\bmit\alpha_i$ and $\bmit\alpha_j$ are in $\{ {\bmit\alpha}\}$ then
${\bmit\alpha_i} + {\bmit\alpha_j}$ must be also.
Thus,
$\{ {\bmit\alpha}\}$ and $\{ {\bmit\beta}\}$ can be defined by choice of
a set of basis vectors $\{{\bmit V}_i\}$

One of the non-contractible
loops (the $\bmit\alpha$-loop by choice) may be regarded as space-like,
and the other loop (the $\bmit\beta$-loop) as time-like.
Each  ${\bmit\alpha}$ corresponds to a set of states (a sector)
in the model that are excitations of the vacuum by
modes of the fermions $\{\psi_n\}$ at frequencies
proportional to $\alpha_n$.
The boundary vectors $\bmit\beta$
contribute a set of GSO projections that act on the states in each sector.
Which physical states survive in a given $\bmit\alpha$ sector is a function
of the phase coefficients
$\{C({{\bmit\alpha}\atop{\bmit\beta}})\}$ (or equivalently
of the
$\{C({{\bmit\alpha}\atop{\bmit V}_i })\}$).

The gauge group in a model
depend on the states in the sectors. In a giving sector $\bmit\alpha$
each complex Weyl fermion $\psi_{n,m}$
carries a U(1) charge, $Q(\psi_{n,m})$, related to its boundary condition:
$$Q(\psi_{n,m})= \alpha_{n,m}/2 + N(\psi_{n,m}) \,\, .\eq{qcharge}$$
$N$ is the fermion number operator, with eigen values of
$\{0,1,-1\}$ for antiperiodic fermions and $\{0,-1\}$ for periodic
fermions.
Hence, for antiperiodic fermions $Q(\psi_{n,m})$ has possible values of
$\{0,\pm 1\}$,
and periodic has  values of $\{\pm\half\}$.

Together, the charges of all states in all sectors form a
lattice upon which the roots and weights of an algebra
can be embedded.
Ref.~\pr{lewellen} demonstrated that SO(10)$_2$ can enter into
a string model in this manner.
In free fermionic models the length-squared of the simple roots of
SO(10) is normalized to two. Increasing the level from one to $K$
has the effect of decreasing the length-squared of the roots by
a factor of 1/K.
Hence, the SO(10)$_2$ simple roots
can be represented on the charge lattice of six complex fermions as
$${(0,0,0,1,0,0),\,\,
(\half,-\half,\half,-\half,0,0),\,\, (0,0,1,0,0,0)
\atop
(0,\half,-\half,0,\epsilon_1,\epsilon_2),\,\,
(0,\half,-\half,0,-\epsilon_1,-\epsilon_2)\,\, .}\eq{roots}$$
Either $\epsilon_1=\half=\epsilon_2$ or
$\epsilon_1=\half=-\epsilon_2$,
depending, respectively, on whether the fifth component
(H$_5$) of the SO(10) CSA is embedded as the
sum or difference of the last two U(1) factors. Since this embedding
requires six complex fermions there is an additional orthogonal
U(1) (denoted U(1)$_{\rm X}$)  algebra embedding present. When
H$_5$ is the sum (difference) of the last two factors,
U(1)$_{\rm X}$ is the difference (sum).
{}From the root embedding it is quite
easy to determine the transformation matrix needed to convert
the dynkin weights of the other representations into charges on
the lattice.
Recall, however,
unitarity allows massless representations only up to the 54.
In the next section I discuss initial findings of my search and
classification free fermionic SO(10)$_2$ models containing
adjoint Higgs and exactly three chiral generations of 16's.
I present the simplest possible of these models.

\vskip .3cm
\no{\bf 4. }{\bf SO(10)} Level-Two {\bf Models}
\vskip .2cm
\sectionnum=4\equationnum=0

Three generation SO(10)$_2$ models require at least ten basis
vectors (BVs).  The first nine vectors of Table 1.~form the core of these
models.
The set of
basis vectors $\{ {\bmit V}0, {\bmit V}1, \dots , {\bmit V}5 \}$, along
with a slightly differing   ${\bmit V}8$ were first introduced in the
SO(10)$_2$ model of ref.~\pr{lewellen}.)
The presence of ${\bmit V}0$
is dictated by modular invariance and is found
in all consistent free fermionic models.
${\bmit V}0 + {\bmit V}0 \equiv 0$ generates
a totally antiperiodic boundary vector (all 0's)
from which arise the graviton, dilaton, and antisymmetric tensor,
along with the Cartan subalgebra of the gauge group. Supersymmetry
requires the presence of  ${\bmit V}1$.
which generates the corresponding massless gravitinos. Similarly,
the sector ${\bmit V}i +{\bmit V}1$ produces the superpartners of the
states in any sector ${\bmit V}i$.
The next three sectors are the generators of the
SO(10)$_2$ gauge group and the states associated with
all combinations of BVs in the set
$\{  0, {\bmit V}2, {\bmit V}3, {\bmit V}4\}$ (denoted by
$\{ {\rm gg}\}$)
form the charge
lattice embedding of SO(10)$_2$ discussed  in section {\bf 3}
using the first six (12)
RM complex (real) fermions. The next 16 RM
fermions are the unpairable real fermions (URFs) that contribute
to the central charge of SO(10) without increasing the rank of the group.
The choice of periodicities for these is fixed (mod physically
equivalent reorderings of the fermions).
Masslessness of the gauge bosons requires that
exactly eight of the URFs in each of ${\bmit V}2$, ${\bmit V}3$, and
${\bmit V}4$ must be periodic,
with four periodic URFs common to any two of these
three BVs, and exactly two periodic fermions common to all three.
Together the first five sectors produce $N=4$ SUSY with the observable
SO(10)$_2$ gauge group and a hidden S0(18)$_1$ group (ignoring that the URFs
are not yet unpairable).
Additionally there are 10's, carring U(1)$_{\rm X}$ charges of $\pm 1$,
originating in these gauge sectors.

Each of  the next three BVs, (${\bmit V}5$, ${\bmit V}6$,
and ${\bmit V}7$),
combine with the eight gauge vectors in $\{ {\rm gg}\}$,
to produce one chiral 16 generation apiece (defined as first second
and third respectively).
The resulting three generations
carry their own local U(1) charges, compliments of complex RM fermions
$\bar\psi_{49,50}$, $\bar\psi_{51,52}$, and $\bar\psi_{53,54}$,
respectively. (Due to the symmetry between RM fermions
$\bar\psi_{47}$ and $\bar\psi_{49,50}$ and between
$\bar\psi_{48}$ and $\bar\psi_{51,52}$
the first and second generation
U(1)'s are enhanced to SU(2)$_2$.)

Chirality of the generations is a result of the GSO projections from two
of the generations acting on the remaining one.
GSO projections from ${\bmit V}5$ and ${\bmit V}6$ also reduce
spacetime SUSY to $N=1$.
 At this stage, eight copies of the first and second and four copies
of the third generation survive the GSO projections.
The existence of generation
U(1) charges reduces the hidden sector gauge group
to SO(10)$_1$.  This demonstrates
the general rule for SO(10)$_2$ models that the rank of the
hidden sector gauge group is never more than five.

In this model
the first generation basis vector is assigned boundary conditions such that
unpairability of the URFs is completed.
This choice leaves no physically significant degrees of freedom (DOF) in
the components of ${\bmit V}5$. On the other hand, there are
several DOF in the BVs for the second and third generations:
around six significant options regarding
which eight URFs are to be periodic in ${\bmit V}6$ and
for each of these, three choices for ${\bmit V}7$. (However, these DOF may
prove to have no effect on the phenomenology of the model,
and may simply demonstrate the high degree of symmetry in free
fermionic strings.)
The periodic URF choices presented below are related to the second and third
generation BVs used in flipped-SU(5) models.\mpr{antoniadis87b}

${\bmit V}8$ is the BV responsible for
the presence of an adjoint Higgs in the model.
Possible variations from ${\bmit V}8$ as given below
are minimal. Modular invariance requires that ${\bmit V}8$  have four
periodic LM fermions in common with two of the generations and none in
common with the third. Choice of which two generations (the first and
second) defines the LM part of the basis vector.
There are
only two significant choices for the ${\bmit V}8$ periodic URFs; these
distinguishing between the first and second generation.
The components of the 45 Higgs are contained in the eight
sectors generated by
${\bmit V}8 + \{ {\rm gg}\}$.
 In this model
${\bmit V}8$ reduces the number of first and second generation copies
in ${\bmit V}6$ and ${\bmit V}7$ down to four also.
{\it However}, unfortunately the new vectors
 ${\bmit V}8 + {\bmit V}5 + \{ {\rm gg}\}$
and ${\bmit V}8 + {\bmit V}6 + \{ {\rm gg}\}$
contribute four new copies of these generations.
The ${\bmit V}8$ GSO either projects out all or keeps all four copies of the
third generation. The appropriate value of
$C({ {\bmit V}7\atop{\bmit V}8})$ is chosen to keep them.
Four copies of the Higgs 45 also survive at this stage.

These first eight sectors  (with their various options) form the basis
of all SO(10) three generation models. If only one more BV,
such as ${\bmit V}9$ below, is to
be added, severe requirements fall upon it.
Since there are four original copies of each generation
and four new copies of the first and second,
this  BV must generate $\Z_4$ GSO projections,
removing all but one copy of each generation from
${\bmit V}5$, ${\bmit V}6$, and ${\bmit V}7$, respectively,
and all copies from
${\bmit V}5+{\bmit V}8$ and ${\bmit V}6+{\bmit V}8$.
(In doing so ${\bmit V}9$ should of course keep the SO(10)$_2$ symmetry.)
This requires some of the components of ${\bmit V}9$ to have
values of $\pm \half$ (denoted simply by $\pm$ in Table 1.).
${\bmit V}9$ should also break the
first generation SU(2) symmetry down to
U(1). (${\bmit V}8$ breaks the
second generation SU(2).)
Last, ${\bmit V}9$ cannot mix with other vectors
to create simultaneous SO(10)$_2$ and hidden sector non-singlet
states that survive GSO projections.
There appear to be several choices,
for ${\bmit V}9$ (along with related $C$-coefficients)
that reduce the copies of generations
in ${\bmit V}5$ through ${\bmit V}7$ to one, allow
one or two Higgs 45's to survive, and project out all the new generation copies
created by the presence of the Higgs.
The ${\bmit V}9$ given below is from the subset of choices that maximizes the
symmetry
between the boundary conditions of the worldsheet fermions
associated with each of the generations.

I have written a computer program to generate a
search for and analysis of the set of phenomenologically unique free fermionic
three
generation SO(10)$_2$ models containing adjoint Higgs. This search is
now underway and will be reported in detail in upcoming ref.~\pr{cleaver94}.
The observable variations found among these models will
correspond to  differing numbers of massless scalar 10's, 16-$\bar 16$ pairs,
and 54's surviving GSO projections.
As I mentioned, the highest dimensional massless SO(10) representation possible
in these models are 54's. Presence of a 54 necessitates a boundary
vector with LM components identical to those of ${\bmit V}8$, but with
all 0's for RM components. Two copies of 54's are required for minimal
SO(10) SUSY-GUTs such as those in \pr{raby}. Perhaps not coincidentally,
two appears to be the maximum number of 54's that might possibly survive the
complete set of GSO projections from ${\bmit V}0$ through ${\bmit V}8$.
How many of these copies (if any) actually survive should be
extremely model dependent. The difficulty of including a BV
responsible for a 54 is identical to that arising from the addition of
${\bmit V}8$: extra copies of the
first and second generations will result unless correct GSO
projections on this new BV are chosen.

\hbox to 1cm{\hfill}

\no Table 1. Set of Basis Vectors for a Free Fermionic SO(10) Level-2
Model With Three Generations and an Adjoint Higgs Scalar

\vskip 7.2truecm

\n {\bf References}
\sectionnumstyle{arabic}

\begin{putreferences}

\ref{abbott84}{L.F.~Abbott and M.B.~Wise, {\it Nucl.~Phys.~}{\bf B244}
(1984) 541.}

\ref{alvarez86}{L.~Alvarez-Gaum\' e, G.~Moore, and C.~Vafa,
{\it Comm.~Math.~Phys.~}{\bf 106} (1986) 1.}

\ref{antoniadis86} {I.~Antoniadis and C.~Bachas, {\it Nucl.~Phys.~}{\bf B278}
 (1986) 343;\\
M.~Hama, M.~Sawamura, and H.~Suzuki, RUP-92-1.}
\ref{li88} {K.~Li and N.~Warner, {\it Phys.~Lett.~}{\bf B211} (1988)
101;\\
A.~Bilal, {\it Phys.~Lett.~}{\bf B226} (1989) 272;\\
G.~Delius, preprint ITP-SB-89-12.}
\ref{antoniadis87}{I.~Antoniadis, C.~Bachas, and C.~Kounnas,
{\it Nucl.~Phys.~}{\bf B289} (1987) 87.}
\ref{antoniadis87b}{I.~Antoniadis, J.~Ellis, J.~Hagelin, and D.V.~Nanopoulos,
{\it Phys.~Lett.~}{\bf B194} (1987) 231.}
\ref{antoniadis88}{I.~Antoniadis and C.~Bachas, {\it Nucl.~Phys.~}{\bf B298}
(1988) 586.}

\ref{ardalan74}{F.~Ardalan and F.~Mansouri, {\it Phys.~Rev.~}{\bf D9} (1974)
3341; {\it Phys.~Rev.~Lett.~}{\bf 56} (1986) 2456;
{\it Phys.~Lett.~}{\bf B176} (1986) 99.}

\ref{argyres91a}{P.~Argyres, A.~LeClair, and S.-H.~Tye,
{\it Phys.~Lett.~}{\bf B235} (1991).}
\ref{argyres91b}{P.~Argyres and S.~-H.~Tye, {\it Phys.~Rev.~Lett.~}{\bf 67}
(1991) 3339.}
\ref{argyres91c}{P.~Argyres, J.~Grochocinski, and S.-H.~Tye, preprint
CLNS 91/1126.}
\ref{argyres91d}{P.~Argyres, K.~Dienes and S.-H.~Tye, preprints CLNS 91/1113;
McGill-91-37.}
\ref{argyres91e} {P.~Argyres, E.~Lyman, and S.-H.~Tye
preprint CLNS 91/1121.}
\ref{argyres91f}{P.~Argyres, J.~Grochocinski, and S.-H.~Tye,
{\it Nucl.~Phys.~}{\bf B367} (1991) 217.}
\reference{argyres93}{P.~Argyres and K.~Dienes,
{\it Phys.~Rev.~Lett.~}{\bf 71} (1993) 819.}

\ref{athanasiu88}{G.~Athanasiu and J.~Atick, preprint IASSNS/HEP-88/46.}

\ref{atick88}{J.~Atick and E.~Witten, {\it Nucl.~Phys.~}{\bf B2 }
(1988) .}

\ref{axenides88}{M.~Axenides, S.~Ellis, and C.~Kounnas,
{\it Phys.~Rev.~}{\bf D37} (1988) 2964.}

\ref{bailin92}{D.~Bailin and A.~Love, {\it Phys.~Lett.} {\bf B292}
(1992) 315.}

\ref{barnsley88}{M.~Barnsley, {\underbar{Fractals Everywhere}} (Academic
Press, Boston, 1988).}

\ref{bouwknegt87}{P.~Bouwknegt and W.~Nahm,
{\it Phys.~Lett.~}{\bf B184} (1987) 359;\\
F.~Bais and P.~Bouwknegt, {\it Nucl.~Phys.~}{\bf B279} (1987) 561;\\
P.~Bouwknegt, Ph.D.~Thesis.}

\ref{bowick89}{M.~Bowick and S.~Giddings, {\it Nucl.~Phys.~}{\bf B325}
(1989) 631.}
\ref{bowick92}{M.~Bowick, SUHEP-4241-522 (1992).}
\ref{bowick93}{M.~Bowick, Private communications.}

\ref{brustein92}{R.~Brustein and P.~Steinhardt, preprint UPR-541T.}

\ref{capelli87} {A.~Cappelli, C.~Itzykson, and
J.~Zuber, {\it Nucl.~Phys.~}{\bf B280 [FS 18]} (1987) 445;
{\it Commun.~Math.~Phys.~}113 (1987) 1.}

\ref{carlitz}{R.~Carlitz, {\it Phys.~Rev.~}{\bf D5} (1972) 3231.}

\ref{candelas85}{P.~Candelas, G.~Horowitz, A.~Strominger, and E.~Witten,
{\it Nucl.~Phys.~}{\bf B258} (1985) 46.}

\ref{cateau92}{H.~Cateau and K.~Sumiyoshi,
{\it Phys.~Rev.~}{\bf D46} (1992) 2366.}

\ref{christe87}{P.~Christe, {\it Phys.~Lett.~}{\bf B188} (1987) 219;
{\it Phys.~Lett.~}{\bf B198} (1987) 215; Ph.D.~thesis (1986).}

\ref{clavelli90}{L.~Clavelli {\it et al.}, {\it Int.~J.~Mod.~Phys.~}{\bf A5}
(1990) 175.}

\ref{cleaver92a}{G.~Cleaver. {\it ``Comments on Fractional Superstrings,''}
To appear in the Proceedings of the International Workshop on String
Theory, Quantum Gravity and the Unification of Fundamental Interactions,
Rome, 1992.}
\ref{cleaver93a}{G.~Cleaver and D.~Lewellen, {\it Phys.~Rev.~}{\bf B300}
(1993) 354.}
\ref{cleaver93b}{G.~Cleaver and P.~Rosenthal, preprint CALT 68/1756.}
\ref{cleaver93c}{G.~Cleaver and P.~Rosenthal, preprint CALT 68/1878.}
 \reference{cleaver94a}{G.~ Cleaver and K.~Dienes,
{\it ``Internal Projection Operators for Fractional Superstrings,''}
OHSTPY-HEP-T-93-022, McGill/93-43. To appear.}
\ref{cleaver94}{G.~Cleaver, {\it ``SO(10) SUSY-GUTS From
Superstrings,''}
OHSTPY-HEP-T-94-008. To appear.}

\ref{clgan}{For recent progress in classification of asymmetric
modular invariants see\\
T.~Gannon, Carleton preprint 92-0407;
hep-th 9408119, 9407055, 9404185,  9402074,\\
G.~Cleaver and D.~Lewellen, {\it Phys.~Lett.~}{\bf B300}
(1993) 354.}

\ref{cornwell89}{J.~F.~Cornwell, {\underbar{Group Theory in Physics}},
{\bf Vol. III}, (Academic Press, London, 1989).}
\ref{dienes92b}{K.~Dienes and S.~-H.~Tye, {\it Nucl.~Phys.~}{\bf B376} (1992)
297.}
\ref{dienes93a}{K.~Dienes, {\it Nucl.~Phys.~}{\bf B413} (1994) 103
(hep-th/9305094).}
\ref{dienes93b}{K.~Dienes, McGill preprint McGill/93-18.
To appear in {\it Nucl.~Phys.~}{\bf B}.}

\ref{deo89a}{N.~Deo, S.~Jain, and C.~Tan, {\it Phys.~Lett.~}{\bf
B220} (1989) 125.}
\ref{deo89b}{N.~Deo, S.~Jain, and C.~Tan, {\it Phys.~Rev.~}{\bf D40}
(1989) 2626.}
\ref{deo92}{N.~Deo, S.~Jain, and C.~Tan, {\it Phys.~Rev.~}{\bf D45}
(1992) 3641.}
\ref{deo90a}{N.~Deo, S.~Jain, and C.-I.~Tan,
in {\underbar{Modern Quantum Field Theory}},
(World Scientific, Bombay, S.~Das {\it et al.} editors, 1990).}

\ref{distler90}{J.~Distler, Z.~Hlousek, and H.~Kawai,
{\it Int.~Jour.~Mod.~Phys.~}{\bf A5} (1990) 1093.}
\ref{distler93}{J.~Distler, private communication.}

\ref{dixon85}{L.~Dixon, J.~Harvey, C.~Vafa and E.~Witten,
 {\it Nucl.~Phys.~}{\bf B261} (1985) 651; {\bf B274} (1986) 285.}
\ref{dixon87}{L.~Dixon, V.~Kaplunovsky, and C.~Vafa,
{\it Nucl.~Phys.~}{\bf B294} (1987) 443.}

\ref{drees90}{W.~Drees, {\underbar{Beyond the Big Bang},}
(Open Court, La Salle, 1990).}

\ref{dreiner89a}{H.~Dreiner, J.~Lopez, D.V.~Nanopoulos, and
D.~Reiss, preprints MAD/TH/89-2; CTP-TAMU-06.}
\ref{dreiner89b}{H.~Dreiner, J.~Lopez, D.V.~Nanopoulos, and
D.~Reiss, {\it Phys.~Lett.~}{\bf B216} (1989) 283.}

\ref{ellis90}{J.~Ellis, J.~Lopez, and D.V.~Nanopoulos,
{\it Phys.~Lett.~}{\bf B245} (1990) 375.}

\ref{fernandez92}{R.~Fern\' andez, J.~Fr\" ohlich, and A.~Sokal,
{\underbar{Random Walks, Critical Phenomena, and Triviality in}}
{\underbar{Quantum Mechanics}}, (Springer-Verlag, 1992).}

\ref{font90}{A.~Font, L.~Ib\'a\~ nez, and F.~Quevedo,
{\it Nucl.~Phys.~}{\bf B345} (1990) 389.}

\ref{frampton88}{P.~Frampton and M.~Ubriaco, {\bf D38} (1988) 1341.}

\ref{francesco87}{P.~di Francesco, H.~Saleur, and J.B.~Zuber,
{\it Nucl.~Phys.~} {\bf B28 [FS19]} (1987) 454.}

\ref{frautschi71}{S.~Frautschi, {\it Phys.~Rev.~}{\bf D3} (1971) 2821.}

\ref{gannon92}{T.~Gfannon, Carleton preprint 92-0407.}

\ref{gasperini91}{M.~Gasperini, N.~S\'anchez, and G.~Veneziano,
{\it Int.~Jour.~Mod.~Phys.~}{\bf A6} (1991) 3853;
{\it Nucl.~Phys.~}{\bf B364} (1991) 365.}

\ref{gepner87}{D.~Gepner and Z.~Qiu, {\it Nucl.~Phys.~}{\bf B285} (1987)
423.}
\ref{gepner87b}{D.~Gepner, {\it Phys.~Lett.~}{\bf B199} (1987) 380.}
\ref{gepner88a}{D.~Gepner, {\it Nucl.~Phys.~}{\bf B296} (1988) 757.}

\ref{ginsparg88}{P.~Ginsparg, {\it Nucl.~Phys.~}{\bf B295 [FS211]}
(1988) 153.}
\ref{ginsparg89}{P.~Ginsparg, in \underbar{Fields, Strings and Critical
Phenomena}, (Elsevier Science Publishers, E.~Br\' ezin and
J.~Zinn-Justin editors, 1989).}

\ref{gross84}{D.~Gross, {\it Phys.~Lett.~}{\bf B138} (1984) 185.}

\ref{green53} {H.~S.~Green, {\it Phys.~Rev.~}{\bf 90} (1953) 270.}

\ref{hagedorn68}{R.~Hagedorn, {\it Nuovo Cim.~}{\bf A56} (1968) 1027.}

\ref{kac80}{V.~Ka\v c, {\it Adv.~Math.~}{\bf 35} (1980) 264;\\
V.~Ka\v c and D.~Peterson, {\it Bull.~AMS} {\bf 3} (1980) 1057;
{\it Adv.~Math.~}{\bf 53} (1984) 125.}
\ref{kac83}{V.~Ka\v c, {\underbar{Infinite Dimensional Lie Algebras}},
(Birkh\" auser, Boston, 1983);\\
V.~Ka\v c editor, {\underbar{Infinite Dimensional Lie Algebras and Groups}},
(World Scientific, Singapore, 1989).}

\ref{kaku91}{M.~Kaku, \underbar{Strings, Conformal Fields and Topology},
(Springer-Verlag, New York, 1991).}

\ref{kawai87a} {H.~Kawai, D.~ Lewellen, and S.-H.~Tye,
{\it Nucl.~Phys.~}{\bf B288} (1987) 1.}
\ref{kawai87b} {H.~Kawai, D.~Lewellen, J.A.~Schwartz,
and S.-H.~Tye, {\it Nucl.~Phys.~}{\bf B299} (1988) 431.}

\ref{kazakov85}{V.~Kazakov, I.~Kostov, and A.~Migdal,
{\it Phys.~Lett.~}{\bf B157} (1985) 295.}

\ref{khuri92}{R.~Khuri, CTP/TAMU-80/1992; CTP/TAMU-10/1993.}

\ref{kikkawa84}{K.~Kikkawa and M.~Yamasaki, {\it Phys.~Lett.~}{\bfB149}
(1984) 357.}

\ref{kiritsis88}{E.B.~Kiritsis, {\it Phys.~Lett.~}{\bf B217} (1988) 427.}

\ref{langacker92}{P.~Langacker, preprint UPR-0512-T (1992).}

\ref{leblanc88}{Y.~Leblanc, {\it Phys.~Rev.}{\bf D38} (1988) 38.}

\ref{lewellen87}{H.~Kawai, D.~Lewellen, and S.-H.`Tye,
{\it Nucl.~Phys.~}{\bf B288} (1987) 1.}
\ref{lewellen}{D.~C.~Lewellen, {\it Nucl.~Phys.~}{\bf B337} (1990) 61.}

\ref{lizzi90}{F.~Lizzi and I.~Senda, {\it Phys.~Lett.~}{\bf B244}
(1990) 27.}
\ref{lizzi91}{F.~Lizzi and I.~Senda, {\it Nucl.~Phys.~}{\bf B359}
(1991) 441.}

\ref{lust89}{D.~L\" ust and S.~Theisen,
{\underbar{Lectures on String Theory,}} (Springer-Verlag, Berlin, 1989).}

\ref{maggiore93}{M.~Maggiore, preprint IFUP-TH 3/93.}

\ref{mansouri87} {F.~Mansouri and X.~Wu, {\it Mod.~Phys.~Lett.~}{\bf A2}
(1987) 215; {\it Phys.~Lett.~}{\bf B203} (1988) 417;
{\it J.~Math.~Phys.~}{\bf 30} (1989) 892;\\
A. Bhattacharyya {\it et al.,} {\it Mod.~Phys.~Lett.~}{\bf A4} (1989)
1121; {\it Phys.~Lett.~}{\bf B224} (1989) 384.}

\ref{narain86} {K.~S.~Narain, {\it Phys.~Lett.~}{\bf B169} (1986) 41.}
\ref{narain87} {K.~S.~Narain, M.H.~Sarmadi, and C.~Vafa,
{\it Nucl.~Phys.~}{\bf B288} (1987) 551.}

\ref{obrien87}{K.~O'Brien and C.~Tan, {\it Phys.~Rev.~}{\bf D36} (1987)
1184.}

\ref{parisi79}{G.~Parisi, {\it Phys.~Lett.~}{\bf B81} (1979) 357.}

\ref{polchinski88}{J.~Polchinski, {\it Phys.~Lett.~}{\bf B209} (1988)
252.}
\ref{polchinski93}{J.~Polchinski, Private communications.}

\ref{pope92}{C.~Pope, preprint CTP TAMU-30/92  (1992).}

\ref{raby}{G.~Anderson, S.~Raby, S.~Dimopoulos, L.J.~Hall, and
G.D.~Starkman,
{\it Phys.~Rev.~}{\bf D49} (1994)
3660;\\
 S.~Raby, talk presented at IFT Workshop on Yukawa Couplings,
Gainsville, Florida, February 1993, hep-th 9406333.}

\ref{raiten91}{E.~Raiten, Thesis, (1991).}

\ref{roberts92}{P.~Roberts and H.~Terao, {\it Int.~J.~Mod.~Phys.~}{\bf A7}
(1992) 2207;\\
P.~Roberts, {\it Phys.~Lett.~}{\bf B244} (1990) 429.}

\ref{sakai86}{N.~Sakaii and I.~Senda, {\it Prog.~Theo.~Phys.~}
{\bf 75} (1986) 692.}

\ref{salomonson86}{P.~Salomonson and B.-S.~Skagerstam, {\it
Nucl.~Phys.~}{\bf B268} (1986) 349.}

\ref{schellekens89} {A.~N.~Schellekens and S.~Yankielowicz,
{\it Nucl.~Phys.~}{\bf B327} (1989) 3;\\
A.~N.~Schellekens, {\it Phys.~Lett.~}{\bf 244} (1990) 255;\\
B.~Gato-Rivera and A.~N.~Schellekens, {\it Nucl.~Phys.~}{\bf B353} (1991)
519; {\it Commun.~Math.}
{\it Phys.~}145 (1992) 85.}
\ref{schellekens89b}{B.~Schellekens, ed. \underbar{Superstring Construction},
 (North-Holland Physics, Amsterdam, 1989).}
\ref{schellekens89c}{B.~Schellekens, CERN-TH-5515/89.}

\ref{schwarz87}{M.~Green, J.~Schwarz, and E.~Witten,
\underbar{Superstring Theory}, {\bf Vols. I \& II},
(Cambridge University Press, New York, 1987).}

\ref{turok87a}{D.~Mitchell and N.~Turok, {\it Nucl.~Phys.~}{\bf B294}
(1987) 1138.}
\ref{turok87b}{N.~Turok, Fermilab 87/215-A (1987).}

\ref{verlinde88}{E.~Verlinde, {\it Nucl.~Phys.~}{\bf B300}
(1988) 360.}

\ref{warner90}{N.~Warner, {\it Commu.~Math.~Phys.~}{\bf 130} (1990) 205.}

\ref{wilczek90} {F.~Wilczek, ed. \underbar {Fractional Statistics and Anyon
Superconductivity}, (World Scientific, Singaore, 1990) 11-16.}

\ref{witten92}{E.~Witten, preprint IASSNS-HEP-93-3.}

\ref{vafa1}{R.~Brandenberger and C.~Vafa, {\it Nucl.~Phys.}
{\bf B316} (1989) 391.}
\ref{vafa2}{A.A.~Tseytlin and C.~Vafa, {\it Nucl.~Phys.}
{\bf B372} (1992) 443.}

\ref{zamol87}{A.~Zamolodchikov and V.~Fateev, {\it Sov.~Phys.~}JETP
{\bf 62} (1985)  215; {\it Teor.~}{\it Mat.}
{\it Phys.~}{\bf 71} (1987) 163.}

\end{putreferences}

\bye